\documentclass[aps,twocolumn,showpacs,groupedaddress,nofootinbib]{revtex4-1}%\documentclass[11pt]{article}
\usepackage{graphicx}
\usepackage{amsmath}
\usepackage{amssymb}
\usepackage{axodraw4j}
\usepackage{pstricks}
\usepackage{color}

\def\dblone{\hbox{$1\hskip -1.2pt\vrule depth 0pt height 1.6ex width 0.7pt
                  \vrule depth 0pt height 0.3pt width 0.12em$}}
\newcommand{\Nu}{\nu}

\begin{document}
\title{The non-linear Glasma}
\author{J\"{u}rgen Berges$^{1,2}$, S\"{o}ren Schlichting$^{1,3}$}
\affiliation{\vspace{0.2cm}
$^1$Institut f{\"u}r Theoretische Physik\index{\footnote{}},
Universit{\"a}t Heidelberg\\
Philosophenweg 16, 69120 Heidelberg
\vspace{0.1cm}\\
$^2$ExtreMe Matter Institute EMMI,\\
GSI Helmholtzzentrum\\ 
Planckstr.~1, 64291~Darmstadt
\vspace{0.1cm} \\
$^3$Theoriezentrum, Institut f{\"u}r Kernphysik\\
Technische Universit{\"a}t Darmstadt\\
Schlossgartenstr.~9, 64289 Darmstadt}

\begin{abstract}
We study the evolution of quantum fluctuations in the Glasma created immediately after the collision of heavy nuclei. It is shown how the presence of instabilities leads to an enhancement of non-linear interactions among initially small fluctuations. The non-linear dynamics leads to an enhanced growth of fluctuations in a large momentum region exceeding by far the originally unstable band.  We investigate the dependence on the coupling constant at weak coupling using classical statistical lattice simulations for $SU(2)$ gauge theory and show how these non-linearities can be analytically understood within the framework of two-particle irreducible (2PI) effective action techniques. The dependence on the coupling constant is only logarithmic in accordance with analytic expectations. Concerning the isotropization of bulk quantities, our results indicate that the system exhibits an order-one anisotropy on parametrically large time scales. Despite this fact, we find that gauge invariant pressure correlation functions seem to exhibit a power law behavior characteristic for wave turbulence.
\end{abstract}

\maketitle

\section{Introduction}
The great interest in relativistic heavy ion collision experiments is to a large part driven by its possibility to explore the properties of deconfined strongly interacting matter described by quantum chromodynamics (QCD). The past decades have revealed remarkable properties of the quark-gluon plasma, probably most strikingly its behavior similar to an ideal fluid \cite{FLUID1,FLUID2}. However these properties are not directly accessible experimentally as they are encoded in the final particle spectra measured by the detectors at RHIC and the LHC \cite{EXP1,EXP2,EXP3,EXP4,EXP5}. Consequently the extraction of medium properties crucially depends on theoretical input, such as the time when the plasma thermalizes locally, which has to be calculated within an ab-initio approach. The non-equilibrium dynamics of high energy nuclear collision poses a challenging problem in the underlying theory of QCD, which in practice can only be addressed with suitable approximations. In this context a field theoretical framework known as the 'color glass condensate' has been developed, which provides a real time ab-initio description of nuclear collisions at high energies \cite{CGCReview,CGCNN,CGC:IC}. While several properties of the initial state right after the collision can be explored within this approach \cite{CGC2DKrasnitz,CGC2DLappi,CGC2DBlaizot,CGC2DMcLerran,CGC2DSchenke}, present studies have not yet been able to explain the thermalization mechanism \cite{GlasmaRV,GlasmaGelis}. Here quantum fluctuations may play an important role as they break the longitudinal boost invariance of the system and can be strongly amplified in the presence of plasma instabilities \cite{GlasmaRV,GlasmaGelis,GlasmaFukushima,INST1,INST2,INST3,INST4,INST5,INST6,HL1,HL2,HL3,HL4,HL5,HL6,Strickland,JB:SU2,JB:SU3}.\\
\\
In this paper we investigate the impact of quantum fluctuations in the color glass condensate description of high energy heavy ion collisions. We work at weak coupling, where the presence of plasma instabilities has been established in previous works \cite{GlasmaRV,GlasmaGelis,GlasmaFukushima}, and present results from classical-statistical lattice simulations along with analytic estimates. We investigate in detail the different dynamical stages of the system undergoing an instability and find that in addition to the 'primary' Weibel type instability \cite{GlasmaRV,GlasmaGelis,GlasmaFukushima}, 'secondary' instabilities emerge due to non-linear interactions of unstable modes. This mechanism is very similar to previous observation in non-expanding gauge-theories \cite{JB:SU2,JB:SU3} and cosmological models \cite{JB:PR} and can be naturally understood in the framework of two-particle irreducible (2PI) effective action techniques \cite{JB:review}.\\
\\
This paper is organized as follows: In Sec.~\ref{sec:HIC} we present a short review of the dynamics of nuclear collisions in the color glass condensate (CGC) framework \cite{CGCReview}. We comment in particular on recent developments to include quantum fluctuations within an ab-initio approach \cite{CGCNN,CGC:IC} and show how the discussion in the literature  is related to a classical-statistical treatment. In Sec.~\ref{sec:res} we present results from classical-statistical lattice simulations. We focus on non-linear effects and obtain the relevant growth rates and set-in times of primary and secondary instabilities. We find that our results are rather insensitive to the value of the strong coupling constant $\alpha_s$ as long as $\alpha_s\ll1$. We also investigate the impact of instabilities on bulk properties of the system such as the ratio of longitudinal pressure to energy density. Here our results indicate that the system remains anisotropic on parametrically large time scales. We summarize our results and conclude with Sec.~\ref{sec:conclusion}.

\section{Non-equilibrium dynamics of nuclear collisions}
\label{sec:HIC}
\subsection{High-energy limit and CGC}
\begin{figure}[t]
\includegraphics[width=0.4\textwidth]{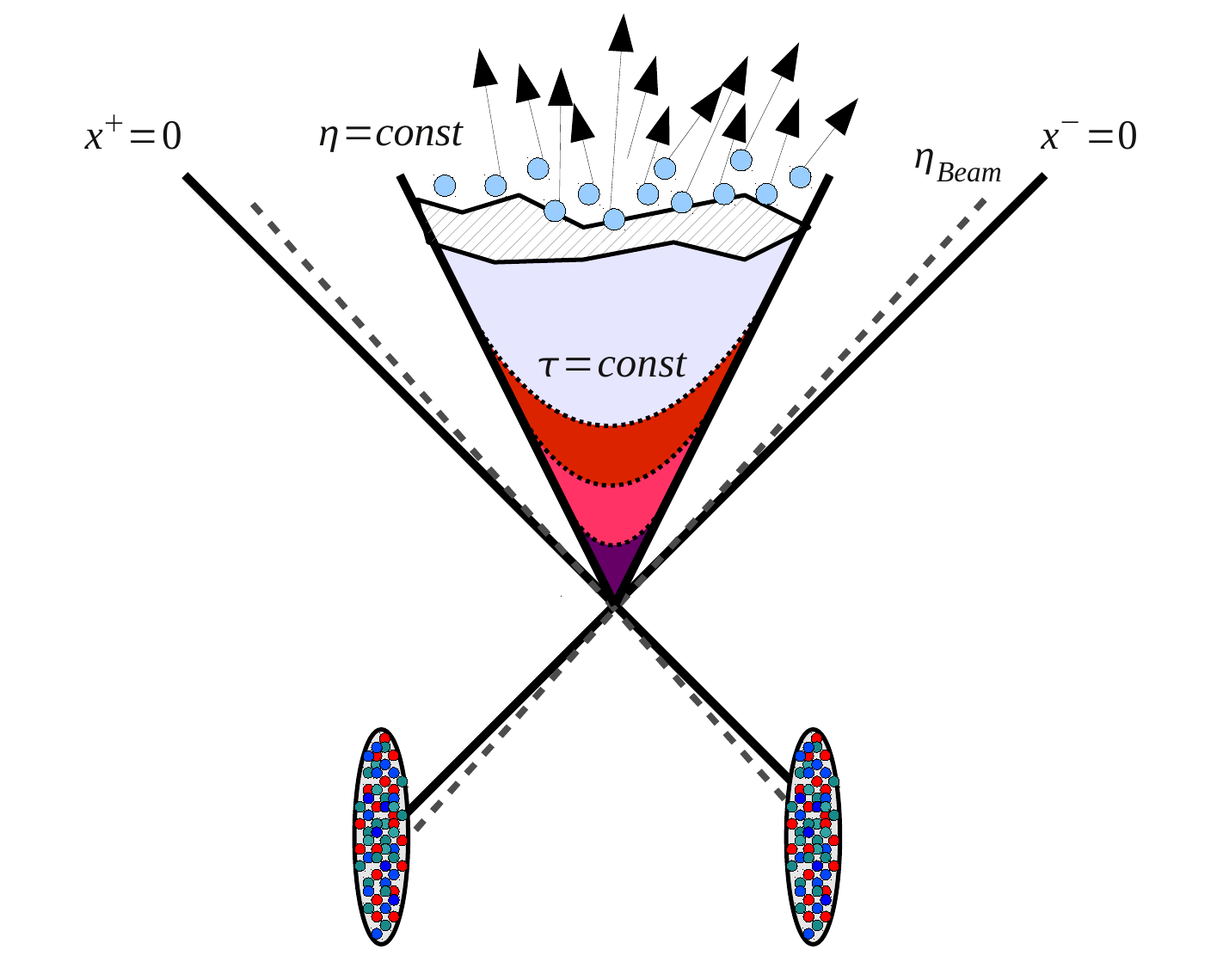}
\caption{\label{fig:cartoon} (color online) Cartoon of the real time evolution of a high energy heavy-ion collision.}
\end{figure}

From a non-equilibrium point of view an ab-initio approach to heavy-ion collisions requires to determine the initial density matrix consisting of two incoming nuclei in the vacuum and subsequently solving the initial value problem in quantum chromodynamics. Though this is beyond the scope of present theoretical methods, one may apply suitable approximations in the high energy and weak coupling limit, which make the problem computationally feasible. This is usually discussed in terms of the light-cone coordinates $x^{\pm}=(t\pm z)/\sqrt{2}$, where at sufficiently high energies the incoming nuclei travel close to the light-cone, which is given by $x^{\pm}=0$. The collision takes place around the time when $x^{+}=x^{-}=0$, where the center of mass of the nuclei coincides and an approximately boost invariant plasma is formed after the collision. The plasma dynamics in the forward light-cone ($x^{\pm}>0$) is usually discussed in terms of the co-moving coordinates
\begin{eqnarray}
\tau=\sqrt{t^2-z^2} \;, \qquad \eta=\text{atanh}(z/t) \;,
\end{eqnarray}
where $\tau$ is the proper time in the longitudinal direction and $\eta$ is the longitudinal rapidity. The metric in these coordinates takes the form $g_{\mu\nu}(x)=\text{diag}(1,-1,-1,-\tau^2)$ and we denote the metric determinant as $g(x)=\text{det}~g_{\mu\nu}(x)$. The dynamics of the collision and the geometry of the coordinates is illustrated in Fig.~\ref{fig:cartoon}. The different colors in the forward light-cone illustrate the dynamics of the longitudinally expanding plasma, which we study in this paper.\\
\\
In the color-glass framework one considers the dynamics of the plasma at mid-rapidity ($\eta\ll\eta_{Beam}$) and the nuclear partons at high rapidities separately. In the eikonal approximation the trajectories of the incoming nuclei are unaffected by the collision, while the dynamics of gluons at mid-rapdity is described by the classical action
\begin{eqnarray}
S[A]=-\frac{1}{4}\int_x \mathcal{F}_{\mu\nu}^a(x)~g^{\mu\alpha}(x)g^{\nu\beta}(x)~\mathcal{F}_{\alpha\beta}^a(x)\;,
\end{eqnarray}
where $\int_x=\int d^4x \sqrt{-g(x)}$ and $\mathcal{F}_{\mu\nu}^a(x)$ denotes the non-abelian field strength tensor
\begin{eqnarray}
\mathcal{F}_{\mu\nu}^a(x)=\partial_\mu A^{\nu}_a(x)-\partial_{\nu}A^{\mu}_a(x)+gf^{abc}A_{\mu}^b(x)A_{\nu}^c(x) \;.
\end{eqnarray}
with Lorentz indices $\mu,\nu$ and color indices $a=1...N_c^2-1$ for $SU(N_c)$ gauge theories. In addition, the gauge field $A_{\mu}^a(x)$ is coupled to an eikonal current $J^{\mu}_a(x)$, which is determined by the properties of the nuclear wavefunction at high rapidities. In practice, the separation in fast and slow degrees of freedom is performed by a renormalization group procedure prescribed by the JIMWLK equations \cite{CGCReview}.  In the high energy limit the eikonal current $J^{\mu}_a(x)$ is given by static color sources on the light-cone and takes the form
\begin{eqnarray}
\label{eq:EikCurr}
J^{\mu}_a(t,x_{\bot},z)=\delta^{\mu+}\varrho^{(1)}_a(x_{\bot})\delta(x^-)+\delta^{\mu-}\varrho^{(2)}_a(x_{\bot})\delta(x^+)\;, \nonumber \\
\end{eqnarray}
where $\delta^{\mu\pm}$ is the Kronecker delta in light-cone coordinates and we denote transverse coordinates as $x_\bot=(x^1,x^2)$. The color charge densities $\varrho^{(1/2)}_a(x_{\bot})$, where the superscript $(1/2)$ labels the different nuclei, contain all further information about the beam energy, nuclear species and impact parameter dependence. At high collider energies, these have been conjectured to exhibit a universal behavior, which is described by the saturation scale $Q_s$ and the value of the strong coupling constant $\alpha_s$ \cite{Saturation1,Saturation2}. In this work we simply adapt the McLerran-Venugopalan (MV) saturation model \cite{MVModel}, where the color charge densities of the nuclei are given by uncorrelated Gaussian configurations
\begin{eqnarray}
\label{MV:sources}
\langle \varrho^{(A)}_a(x_{\bot})\varrho^{(B)}_b(y_{\bot})\rangle=g^2\mu^2~\delta^{AB}~\delta_{ab}~\delta(x_\bot-y_\bot)\;.
\end{eqnarray}
The model parameter $g^2\mu$ is proportional to the physical saturation scale $Q_s$ up to logarithmic corrections and reflects the properties of the saturated wavefunctions of large nuclei \cite{MVModel}. Consequently the current in Eq.~(\ref{eq:EikCurr}) is parametrically large, i.e. formally $\mathcal{O}(1/g)$ in powers of the coupling constant. This makes the problem inherently non-perturbative and we will come back to this aspect when we discuss the impact of quantum fluctuations. In addition to Eq.~(\ref{MV:sources}), we impose a color neutrality constraint on the color charge densities such that the global color charge vanishes separately for each nucleus, i.e.
\begin{eqnarray}
\label{eq:nocolor}
\int d^2x_{\bot}~\varrho^{(1/2)}_a(x_{\bot})=0\quad \forall\quad a.
\end{eqnarray}
By specifying the eikonal current according to Eq.~(\ref{eq:EikCurr}), the longitudinal geometry of the collision has effectively been reduced to the collision of two-dimensional sheets and there is no longer a longitudinal scale inherent to the problem. However quantum fluctuations explicitly break the longitudinal boost invariance of the system and may therefore play an important role in the non-equilibrium dynamics right after the collision \cite{CGC:IC}. Before we turn to a detailed discussion of quantum fluctuations, we will briefly review the classical solution to the particle production process. We will show later, in Sec.~\ref{sec:quantum}, how this solution emerges also in the weak-coupling limit of the quantum field theory, where quantum fluctuations can be handled properly.
\subsection{Classical solution}
\label{sec:classSol}
Neglecting quantum fluctuations for the moment, the absence of a longitudinal scale in Eq.~(\ref{eq:EikCurr}) leads to boost-invariant solutions of the classical Yang-Mills field equations
\begin{eqnarray}
\label{eq:classEOM}
\frac{\delta S[A]}{\delta A_{\mu}^a(x)}=-J^{\mu}_a(x) \;.
\end{eqnarray}
In the classical color glass picture, the strong color-fields right after the collision are entirely determined by the continuity conditions on the light-cone ($x^{\pm}=0$) \cite{CGCclass1,CGCclass2,CGCclass3,CGCclass4}. Adapting the Fock-Schwinger gauge condition ($A_{\tau}=0$), where the classical Yang-Mills action takes the form
\begin{eqnarray}
S[A]&=&\int \tau d\tau~d\eta~d^2x_\bot\left[ \frac{1}{2\tau^2}(\partial_\tau A_{\eta}^a)^2+\frac{1}{2}(\partial_\tau A_i^a)^2\right. \nonumber \\
&&\left.-\frac{1}{2\tau^2} \mathcal{F}_{\eta i}^a \mathcal{F}_{\eta i}^a-\frac{1}{4}\mathcal{F}_{ij}^a\mathcal{F}_{ij}^a\right]\;, 
\end{eqnarray}
($i=1,2$) the initial state right after the collision can be specified at $\tau=0^+$, where the chromo magnetic and electric fields are given by \cite{CGCclass1,CGCclass2,CGCclass3,CGCclass4}
\begin{eqnarray}
\label{eq:IC1}
A_i(x_{\bot})&=&\alpha_i^{(1)}(x_{\bot})+\alpha_i^{(2)}(x_{\bot}) \;, \qquad A_{\eta}=0\;, \\
E_i&=&0 \;, \quad E_{\eta}(x_{\bot})=ig[\alpha_i^{(1)}(x_{\bot})\;,\alpha_i^{(2)}(x_{\bot})]\;. \nonumber
\end{eqnarray}
Here $\alpha_i^{(1/2)}(x_{\bot})$ are pure gauge configurations which describe the Yang-Mills field outside the light-cone. They are related to the nuclear color charge densities by \cite{CGCclass1,CGCclass2,CGCclass3,CGCclass4}
\begin{eqnarray}
\label{eq:alpha}
\alpha^{(N)}_i(x_{\bot})&=&\frac{-i}{g}e^{ig\Lambda^{(N)}(x_{\bot})}\partial_ie^{-ig\Lambda^{(N)}(x_{\bot})} \;, \nonumber \\
\partial_i\partial^{i}\Lambda^{(N)}(x_{\bot})&=&\varrho^{(N)}(x_{\bot})\;,
\end{eqnarray}
and depend on transverse coordinates only. The relations (\ref{eq:IC1}) and (\ref{eq:alpha})
specify the 'Glasma' initial state at $\tau=0^+$ right after the collision. The time evolution in the forward light-cone can be studied numerically by solving the lattice analogue of the classical evolution equations and has been studied extensively \cite{CGC2DKrasnitz,CGC2DLappi,CGC2DBlaizot,CGC2DMcLerran}. However the longitudinal boost invariance of the system is preserved in this classical evolution and leads to an effectively 2+1 dimensional Yang Mills theory coupled to an adjoint scalar field \cite{CGC2DKrasnitz,CGC2DLappi,CGC2DBlaizot,CGC2DMcLerran}. In order to study the full 3+1 dimensional Yang Mills theory it is therefore crucial to include quantum fluctuations, which break the boost-invariance of the system explicitly.
\subsection{Quantum fluctuations}
\label{sec:quantum}
The inclusion of quantum fluctuations has recently attracted great attention due to their expected importance in understanding the thermalization process \cite{CGC:IC,CGCNN}. This concerns in particular the inclusion of vacuum fluctuations in the initial state, which are quantum in origin but evolve classically at sufficiently weak coupling and short enough times \cite{JB:QvsCS}. In contrast to most discussions in the literature \cite{CGC:IC,CGCNN}, our analysis is based on the two particle irreducible (2PI) effective action framework \cite{JB:review}. This formalism has been successfully applied to a variety of similar problems in scalar theories \cite{JB:PR,SS:PR} and gauge theories \cite{JB:SU2,JB:SU3,JB:gauge2PI} and recently been developed for the problem under consideration here \cite{Hatta}. We give a short general introduction to the formalism and show first how the realization proposed in Ref.~\cite{CGC:IC} emerges in this framework. Then we go beyond that discussion and identify sub-leading quantum corrections and describe the non-linear dynamics of instabilities analytically.\\
\\
The quantum evolution equations can be formulated in terms of the expectation values of the gauge field operators $\hat{A}_{\mu}^a(x)$ denoted by
\begin{eqnarray}
\label{eq:2PIfield}
A_{\mu}^a(x)=\langle\hat{A}_\mu^a(x)\rangle\;,
\end{eqnarray}
and the time ordered two-point correlation function
\begin{eqnarray}
G_{\mu\nu}^{ab}(x,y)=\langle \mathcal{T}\hat{A}_{\mu}^a(x) \hat{A}_{\nu}^b(y)\rangle\;,
\end{eqnarray}
The two independent parts of the propagator $G_{\mu\nu}^{ab}(x,y)$ can be expressed in terms of the spectral and statistical two-point correlation functions
\begin{eqnarray}
G_{\mu\nu}^{ab}(x,y)=F_{\mu\nu}^{ab}(x,y)-\frac{i}{2}\text{sgn}(x^0-y^0)\rho^{ab}_{\mu\nu}(x,y) 
\end{eqnarray}
which are associated to the commutator and anti-commutator
\begin{eqnarray}
\rho_{\mu\nu}^{ab}(x,y)&=&i\left< \left[\hat{A}_{\mu}^{a}(x),\hat{A}_{\nu}^{b}(y)\right]  \right>\;, \\
F_{\mu\nu}^{ab}(x,y)&=&\frac{1}{2}\left< \left\{\hat{A}_{\mu}^{a}(x),\hat{A}_{\nu}^{b}(y)\right\}\right>-A_{\mu}^a(x)A_{\nu}^b(y)   \;.
\end{eqnarray}
Here expectation values are given by the trace over the initial vacuum density matrix in the presence of the eikonal currents. The initial density matrix is specified in the remote past ($t_0\to-\infty$), where the background field $A_{\mu}^a$ vanishes and the statistical fluctuations $F_{\mu\nu}^{ab}$ take the vacuum form (see e.g. Ref.~\cite{Hatta}). In contrast, the initial values of the spectral function are entirely determined by the equal time commutation relations, which in temporal gauge ($A_0=0$) read
\begin{eqnarray}
\label{eq:commrel}
\left.\rho_{\mu\nu}^{ab}(x,y)\right|_{x^0=y^0}&=&0\;, \nonumber \\
\left.\partial_{x^0}\rho_{\mu\nu}^{ab}(x,y)\right|_{x^0=y^0}&=&-\delta^{ab}\frac{g_{\mu\nu}}{\sqrt{-g(x)}} \delta(\vec{x}-\vec{y}) \;, \nonumber \\
\left.\partial_{x^0}\partial_{y^0}\rho_{\mu\nu}^{ab}(x,y)\right|_{x^0=y^0}&=&0 \;,
\end{eqnarray}
and are valid at all times.\footnote{Note that Eq.~(\ref{eq:commrel}) is valid also for $(\tau,\eta)$ coordinates, when replacing $x^0=\tau$ and $x^3=\eta$ and imposing the Fock-Schwinger ($A_{\tau}=0$) gauge condition.} The gauge field expectation values in Eq.~(\ref{eq:2PIfield}) correspond to the Glasma background fields, while the spectral and statistical two-point functions contain the quantum fluctuations. The evolution equations for connected one and two-point correlation functions follow from the stationarity of the two particle irreducible (2PI) effective action \cite{Baym,CJT}
\begin{eqnarray}
\Gamma_{\text{2PI}}[A,G]=S[A]&+&\frac{i}{2}\text{tr}\big[\text{ln}G^{-1}\big]+\frac{i}{2}\text{tr}\big[G_0^{-1}[A]G\big] \nonumber \\
&+&\Gamma_2[A,G]+\text{const}\;.
\end{eqnarray}
and form a closed set of coupled evolution equations. The set of equations is given by the evolution equation of the macroscopic field
\begin{eqnarray}
\label{eq:AEOM}
\frac{\delta S[A]}{\delta A_{\mu}^a(x)}=-J^{\mu}_a(x)-\frac{i}{2}\text{tr}\left[\frac{\delta G_0^{-1}[A]}{\delta A_{\mu}^a(x)}G\right]-\frac{\delta \Gamma_2[A,G]}{\delta A_{\mu}^a(x)} \nonumber \\
\end{eqnarray}
and the evolution equations for spectral and statistical two point correlation functions, which can be written as \cite{JB:review}
\begin{figure}[t]
\includegraphics[width=0.4\textwidth]{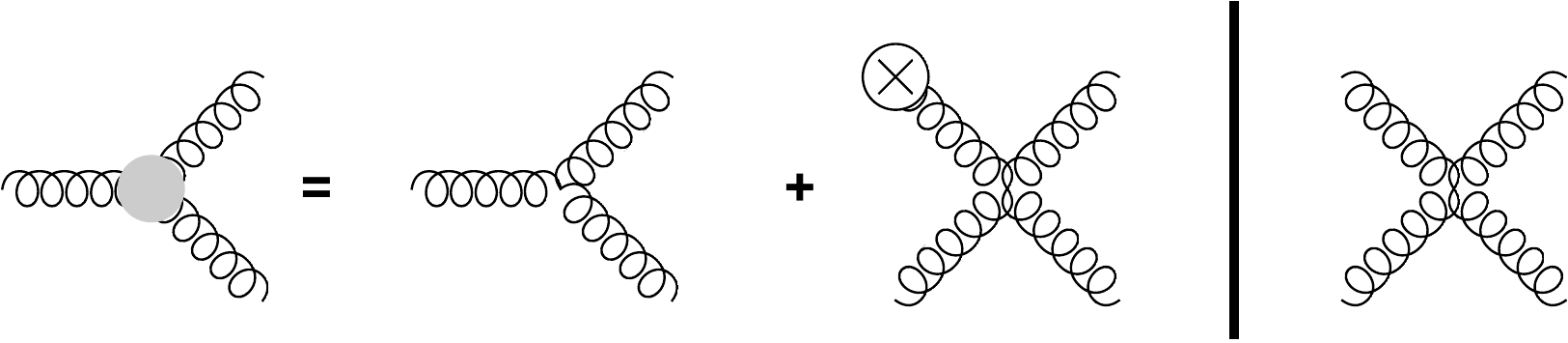}
\caption{\label{fig:Vertices} Vertices in non-abelian gauge theory in the presence of background gauge fields.}
\end{figure} 
\begin{widetext}
\begin{eqnarray}
\label{eq:rhoEOM}
\left[iG_{0,ac}^{-1,\mu\gamma}[x;A]+\Pi^{(0),\mu\gamma}_{ac}(x)\right]\rho_{\gamma\nu}^{cb}(x,y)=&-&\int_{y^0}^{x^0}dz~\Pi^{(\rho),\mu\gamma}_{ac}(x,z)\rho_{\gamma\nu}^{cb}(z,y)\;,\\
\label{eq:FEOM}
\left[iG_{0,ac}^{-1,\mu\gamma}[x;A]+\Pi^{(0),\mu\gamma}_{ac}(x)\right]F_{\gamma\nu}^{cb}(x,y)=&-&\int_{-\infty}^{x^0}dz~\Pi^{(\rho),\mu\gamma}_{ac}(x,z)F_{\gamma\nu}^{cb}(z,y)+\int_{-\infty}^{y^0}dz~\Pi^{(F),\mu\gamma}_{ac}(x,z)\rho_{\gamma\nu}^{cb}(z,y)\;.
\end{eqnarray}
\end{widetext}
Here we  denote $\int_a^b dz=\int_a^b dz^0 \int d^dz \sqrt{-g(z)}$ and $iG_{0,ab}^{-1,\mu\nu}[x;A]$ denotes the free inverse propagator
\begin{eqnarray}
\label{eq:G0inv}
iG_{0,ab}^{-1,\mu\nu}[x;A]&=&~~\gamma^{-1}(x)~D_{\gamma}^{ac}[A]~\gamma(x)~g^{\gamma\alpha}g^{\mu\nu}~D_{\alpha}^{cb}[A] \nonumber \\
&&-\gamma^{-1}(x)~D_{\gamma}^{ac}[A]~\gamma(x)~g^{\gamma\nu}g^{\mu\alpha}~D_{\alpha}^{cb}[A] \nonumber \\
&&-g~f^{abc}~F^{\mu\nu}_c(x)[A]\;,
\end{eqnarray}
with $\gamma(x)=\sqrt{-g(x)}$ and we introduced the (background) covariant derivative 
\begin{eqnarray}
D_{\mu}^{ab}[A]=\partial_{\mu}\delta^{ab}-gf^{abc}A_{\mu}^c\;.
\end{eqnarray}
The non-zero spectral and statistical parts of the self-energy $\Pi^{(\rho/F)}[A,G]$ on the right hand side and the local part  $\Pi^{(0)}[G]$ on the left hand side make the evolution equations non-linear in the fluctuations. In general they contain contributions from the vertices depicted in Fig.~\ref{fig:Vertices}, where in addition to the classical three gluon vertex there is a three gluon vertex associated with the presence of a non-vanishing background field. The explicit expressions for the derivatives on the right hand side of Eq.~(\ref{eq:AEOM}) and the self-energy contributions entering Eqns.~(\ref{eq:rhoEOM}) and (\ref{eq:FEOM}) have been calculated to three loop order ($g^6$) in Ref.~\cite{JB:gauge2PI} and the corresponding expressions in co-moving $(\tau,\eta)$ coordinates can be found in Ref.~\cite{Hatta}. Before we turn to a more detailed discussion of the right hand side contributions of Eqns.~(\ref{eq:AEOM}),~(\ref{eq:rhoEOM}) and (\ref{eq:FEOM}), it is insightful to consider first the leading part in a weak coupling expansion. We will see shortly how this recovers the classical solution for the background field, as discussed in Sec.~\ref{sec:classSol}, while initial state vacuum fluctuations are already included to leading order in terms of the connected two-point correlation functions.\\
\\
In order to isolate the leading contributions one has to take into account the strong external currents $J^{\mu}_a\sim\mathcal{O}(1/g)$, which induce non-perturbatively large background fields $A_{\mu}^a(x)\sim\mathcal{O}(1/g)$. In contrast, the statistical fluctuations $F_{\mu\nu}^{ab}(x,y)$ originate from initial state vacuum and are therefore initally $\mathcal{O}(1)$. The spectral function $\rho_{\mu\nu}^{ab}(x,y)$ has to comply with the equal time commutation relations (\ref{eq:commrel}) and is therefore parametrically $\mathcal{O}(1)$ at any time. Considering only the leading contributions in a weak coupling expansion, the evolution equation (\ref{eq:AEOM}) reduces to its classical form (c.f. Eq.~(\ref{eq:classEOM}))
\begin{eqnarray}
\label{eq:linA}
\frac{\delta S[A]}{\delta A_{\mu}^a(x)}=-J^{\mu}_a(x)\;,
\end{eqnarray}
and the evolution equations for the spectral and statistical two-point correlation functions at leading order read
\begin{eqnarray}
\label{eq:linRho}
iG_{0,ac}^{-1,\mu\gamma}[x;A]~\rho_{\gamma\nu}^{cb}(x,y)&=&0 \;, \\
 iG_{0,ac}^{-1,\mu\gamma}[x;A]~F_{\gamma\nu}^{cb}(x,y)&=&0\;,
\label{eq:linF}
\end{eqnarray}
where sub-leading contributions are suppressed by at least a factor of $g^2$ relative to the leading contribution.
It is important to realize that at this order the evolution of the Glasma background fields decouples from that of the fluctuations, i.e. there is no back-reaction from the fluctuations on the background fields.
Therefore the dynamics of the background fields remains unchanged and one recovers the classical field solutions discussed in Sec.~\ref{sec:classSol}.
In addition the evolution of vacuum fluctuations of the initial state is taken into account by Eqns.~(\ref{eq:linRho}) and (\ref{eq:linF}) to linear order in the fluctuations.
To this order the quantum field theory is known to agree with the classical statistical theory \cite{JB:QvsCS} and Eqns.~(\ref{eq:linRho},~\ref{eq:linF}) can equivalently be obtained by considering the linearized classical evolution equations for small fluctuations \footnote{This can be seen immediately by solving Eq.~(\ref{eq:linF}) in terms of mode functions $F(x,y)=\delta A(x)\delta A(y)$, which then indivdually satisfy the linearized classical evolution equations $iG_0^{-1}[x;A]~\delta A(x)=0\;$.} (see e.g. Ref.~\cite{CGC:IC}).
The linear approximation in Eqns.~(\ref{eq:linRho}) and (\ref{eq:linF}) yields a major simplification to Eqns.~(\ref{eq:rhoEOM}) and (\ref{eq:FEOM}), as one can solve Eqns.~(\ref{eq:linRho}) and (\ref{eq:linF}) independently from the evolution of the background field.
This has been exploited in Ref.~\cite{CGC:IC} to obtain the spectrum of initial fluctuations right after the collision.
In turn, the range of validity of the approximation is limited to the domain where fluctuations remain parametrically small. This is, however, not the case in the forward light-cone ($\tau>0$), where Eqns.~(\ref{eq:linRho}) and (\ref{eq:linF}) exhibit plasma instabilities associated to exponential growth of statistical fluctuations \cite{GlasmaRV,GlasmaGelis}
\begin{eqnarray}
\label{eq:expgrowth}
F_{\mu\nu}^{ab}(\tau,\tau',x_T,y_T,\Nu)\propto\exp[\Gamma(\Nu)(\sqrt{g^2\mu\tau}+\sqrt{g^2\mu\tau'})]\;, \nonumber \\
\end{eqnarray}
for characteristic modes, where $\Nu$ is the Fourier coefficient with respect to relative rapidity according to
\begin{eqnarray}
F_{\mu\nu}^{ab}(x,y)=\int \frac{d\nu}{2\pi} F_{\mu\nu}^{ab}(x,y,\Nu) e^{i\Nu(\eta_x-\eta_y)}\;.
\end{eqnarray}
In this regime fluctuations can become parametrically large and strongly modify the naive power counting. 
Since the approximation underlying Eqns.~(\ref{eq:linRho}) and (\ref{eq:linF}) is not energy conserving, one encounters exponential divergences when stressing Eqns.~(\ref{eq:linRho},~\ref{eq:linF}) beyond their range of validity \cite{Gelis:Div}. In this regime it is crucial to include higher order self-energy corrections, which naturally cure the divergencies. The associated power counting is discussed in Sec.~\ref{sec:pc}, where we identify the diagrammatic contributions, which contain the first relevant corrections and analyze their impact on the dynamics of the instability. A systematic way to include self-energy corrections to all orders was outlined in Ref.~\cite{CGC:IC} and will be discussed in more detail in Sec.~\ref{sec:classStat} on the classical-statistical approximation.
\subsection{Dynamical power counting and non-linear dynamics}
\label{sec:pc}
\begin{table}[t]
\centering
\begin{tabular}{||c||}
\hline
Dynamical power counting \\  \hline \hline
 $\Pi^{(F)}$: one loop \\
\includegraphics[height=1cm]{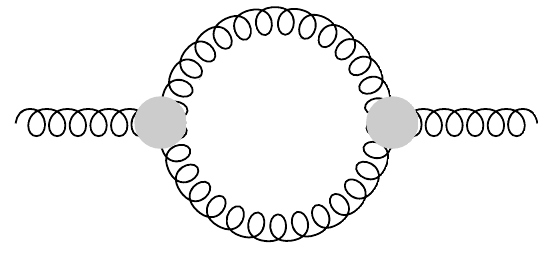} \\ 
 $g^2~F^2\;,~\beta=1$\\ \hline \hline \hline
 $\Pi^{(F)}$: two loop  \\
\includegraphics[height=1cm]{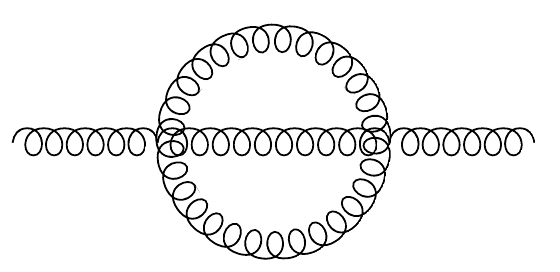} \includegraphics[height=1cm]{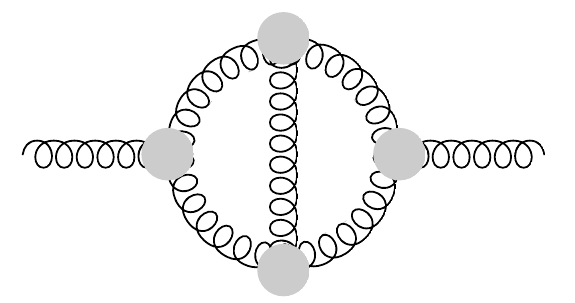} \\ 
\includegraphics[height=1cm]{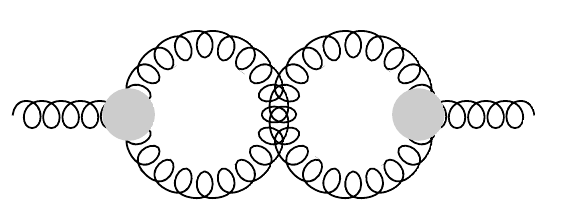} \includegraphics[height=1cm]{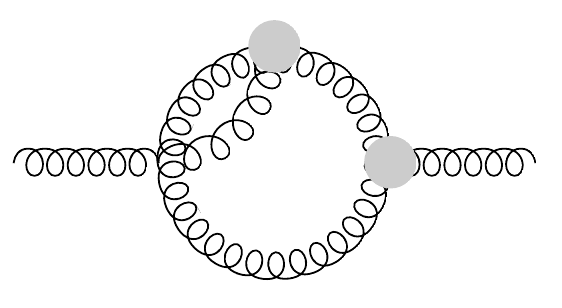} \\
 $g^4~F^3(\rho)\;,~\beta=4/3$\\ \hline \hline \hline
$\Pi^{(0)}$ (local)  \\
\includegraphics[height=1cm]{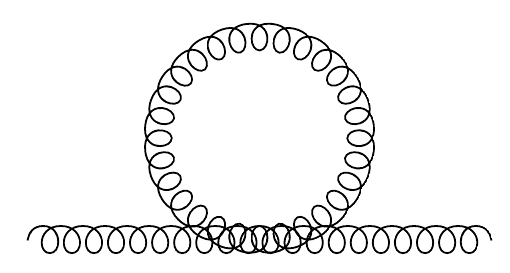}  \\
 $g^2~F\;,~\beta=2$\\ \hline \hline
\end{tabular}
\caption{\label{tab:pc}Self energy diagrams to two loop order ($g^4$). The value $\beta$ corresponds to the classification in the dynamical power-counting scheme. The one-loop diagram in the top panel yields the first relevant correction, while diagrams with higher values of $\beta$ become important at later times.}
\end{table}
In order to go beyond a fixed-order coupling expansion of the 2PI effective action, one has to develop a power counting scheme which takes into account not only the suppression by the small coupling constant but also the enhancement due to parametrically large fluctuations in the presence of non-equilibrium instabilities. This has been worked out in detail for scalar theories \cite{JB:PR,SS:PR} and applies in a similar way to gauge theories \cite{JB:SU2,JB:SU3}. In this power counting scheme self-energy corrections are classified according to powers of the coupling constant $g$ as well as powers of the background field $A_{\mu}^a(x)$ and the statistical fluctuations $F_{\mu\nu}^{ab}(x,y)$. For a generic self-energy contribution containing powers $g^nF^mA^{l}\rho^k$, the integers $n,m$ and $l$ yield the suppression factor from the coupling constant ($n$) as well as the enhancement due to a parametrically large background field ($l$) and parametrically large fluctuations ($m$). The 'weight' of the spectral function ($k$) remains parametrically of order one at all times as encoded in the equal-time commutation relations. For parametrically large macroscopic fields $A_{\mu}^a(x)\sim1/g$ one expects a sizable self-energy correction once fluctuations have grown as large as $F_{\mu\nu}^{ab}(x,y)\sim1/g^{(n-l)/m}$ for characteristic modes. The hierarchy emerging from a classification of diagrams in terms of $\beta=(n-l)/m$ is shown in Tab.~\ref{tab:pc}. The one loop diagram shown in the upper panel contains two three gluon vertices, which give rise to a suppression factor $g^2$ ($(n-l)=2$). On the other hand the diagram is enhanced by two statistical propagators in the loop ($m=2$) and we can classify the overall contribution as $\mathcal{O}(g^2F^2)$. Similarly one can analyze the two loop diagrams and the tadpole diagram also shown in Tab.~\ref{tab:pc}. The tadpole diagram contains a suppression factor $g^2$ from the four gluon vertex ($n=2,~l=0$) and one statistical propagator ($m=1$), such that the overall contribution can be classified as $\mathcal{O}(g^2F)$. The two loop diagrams are of order $g^4$ ($n-l=4$) in the coupling constant and contain at most three statistical propagators ($m=3$). The overall contribution is thus $\mathcal{O}(g^4F^3)$ in the dynamical power counting. The classification in terms of $\beta=(n-l)/m$ shows that the only diagram yielding a contribution to $\beta=1$ is the one-loop diagram in the upper panel of Tab.~\ref{tab:pc}. The leading contribution of higher order loop diagrams can be classified as $\beta=2L/(L+1)$, where $L\geq1$ is the number of loops. The tadpole diagram yields $\beta=2$. Finally there are self-energy contributions which contain powers of the spectral function instead of statistical propagators. These give rise to even higher values of $\beta\geq2$ and are therefore suppressed at sufficiently weak coupling.\\
\\
This hierarchy of diagrams has important consequences for the dynamics of systems undergoing an instability, where initially small fluctuations grow exponentially in time. At early times statistical fluctuations are small and the system is accurately described by the set of equations (\ref{eq:linA}, \ref{eq:linRho}, \ref{eq:linF}), which give rise to the linear instability regime \cite{GlasmaRV,GlasmaGelis}. At later times, when statistical fluctuations have grown larger, self-energy corrections become important and alter the dynamics of the system. In this regime self-energy contributions with smaller values of $\beta$ become important at an earlier stage as compared to contributions with higher values of $\beta$, as each diagram requires $F_{\mu\nu}^{ab}(x,y)\sim g^{-\beta}$ to yield a significant contribution. The one-loop diagram in the top panel of Tab.~\ref{tab:pc} is of order $\mathcal{O}(g^2F^2)$ in the dynamical power counting. The contribution of this diagram becomes of $\mathcal{O}(1)$ as soon as statistical fluctuations have grown as large as $\mathcal{O}(1/g)$, while all higher order self-energy corrections are still suppressed by at least a fractional power of the coupling constant. Accordingly there is a regime where the one loop diagram displayed in the top panel of Tab.~\ref{tab:pc} yields the only relevant correction.\\
\\
In order to investigate the impact of the one-loop correction in more detail, we will in the following neglect all contributions to the memory integrals in Eqns.~(\ref{eq:rhoEOM},~\ref{eq:FEOM}) which originate from outside the forward light cone. This assumption is justified at weak coupling, where fluctuations are sufficiently small until the time when they exhibit exponential growth due to the presence of the instability. We can then switch to a discussion in $(\tau,\eta)$ coordinates and employ the Fock-Schwinger ($A_\tau=0$) gauge condition in the following. The self-energy contributions from the one-loop diagram in Tab.~\ref{tab:pc} (top panel) then take the form \cite{Hatta}
\begin{widetext}
\begin{eqnarray}
\label{eq:piF1loop}
\Pi^{(F),\mu\mu'}_{\textbf{-o-},aa'}(x,y,\Nu)&=& -\frac{g^2}{2}\int_{\Nu'} \overrightarrow{V}_{x,abc}^{\mu\nu\gamma}\left[F_{\nu\nu'}^{bb'}(x,y,\Nu')F_{\gamma\gamma'}^{cc'}(x,y,\Nu-\Nu')-\frac{1}{4}\rho_{\nu\nu'}^{bb'}(x,y,\Nu')\rho_{\gamma\gamma'}^{cc'}(x,y,\Nu-\Nu')\right]\overleftarrow{V}_{y,a'b'c'}^{\mu'\nu'\gamma'}\;,\\
\Pi^{(\rho),\mu\mu'}_{\textbf{-o-},aa'}(x,y,\Nu)&=& -\frac{g^2}{2}\int_{\Nu'} \overrightarrow{V}_{x,abc}^{\mu\nu\gamma}\left[F_{\nu\nu'}^{bb'}(x,y,\Nu')\rho_{\gamma\gamma'}^{cc'}(x,y,\Nu-\Nu')+\rho_{\nu\nu'}^{bb'}(x,y,\Nu')F_{\gamma\gamma'}^{cc'}(x,y,\Nu-\Nu')\right]\overleftarrow{V}_{y,a'b'c'}^{\mu'\nu'\gamma'}\;,
\label{eq:piRho1loop}
\end{eqnarray}
\end{widetext}
where $x=(\tau_x,x_\bot)$ collectively labels the proper time and transverse coordinates, $\Nu$ is the Fourier coefficient with respect to relative rapidity and Lorentz indices take the values $\mu=1,2,\eta$. The three gluon vertices in Eqns.~(\ref{eq:piF1loop}) and (\ref{eq:piRho1loop}) take the form
\begin{eqnarray}
V_{x,abc}^{\mu\nu\gamma}&=&V_{x,abc}^{\textbf{0},\mu\nu\gamma}+V_{x,abc}^{A,\mu\nu\gamma}\;,
\end{eqnarray}
where in addition to the classical three gluon vertex there is a contribution from the background field. The classical three gluon vertex can be written as
\begin{eqnarray}
V_{x,abc}^{\textbf{0},\mu\nu\gamma}&=&f^{abc}\left[g^{\mu\nu}(x)(-\tilde{\partial}_c^{\gamma,x}-2\tilde{\partial}_b^{\gamma,x})\right. \\
&&+\left.g^{\nu\gamma}(x)(\tilde{\partial}_b^{\mu,x}-\tilde{\partial}_c^{\mu,x})+g^{\mu\gamma}(x)(2\tilde{\partial}_c^{\nu,x}+\tilde{\partial}_b^{\nu,x})\right]\nonumber
\end{eqnarray}
(no summation over $b$ and $c$), where the derivative operator $\tilde{\partial}_a^{\mu,x}=(\partial_{\tau_x},-\partial_{x_\bot},-i\tau_x^{-2}\Nu)$ only acts on the propagator with color index $a$, and the contribution from the background field is given by
\begin{eqnarray}
V^{A,\mu\nu\gamma}_{x,abc} &=&-g \left[~g^{\mu\nu}(x)C_{ab,cd}~A^{\gamma}_d(x)\right. \\
&&\left.\quad+~g^{\nu\gamma}C_{bc,ad}~A^{\mu}_d(x)+g^{\mu\gamma}C_{ca,bd}~A^{\nu}_d(x)\right] \nonumber
\label{eq:background}
\end{eqnarray}
with
\begin{eqnarray}
C_{ab,cd}=f^{ace}f^{bde}+f^{ade}f^{bce} \, .
\end{eqnarray}
In order to investigate the impact of the above self-energy corrections, we first note that the $\rho\rho$ term in Eq.~(\ref{eq:piF1loop}) is a genuine quantum correction, which is absent in the classical statistical theory \cite{JB:QvsCS}. On the other hand statistical fluctuations grow exponentially  according to Eq.~(\ref{eq:expgrowth}) for unstable modes, such that the dominant contribution in Eq.~(\ref{eq:piF1loop}) arises from the classical ($FF$) contributions and one can safely neglect the sub-leading quantum corrections. The right hand side of the evolution equations (\ref{eq:FEOM}) then receive exponentially enhanced contributions from the self-energies (\ref{eq:piF1loop}) and (\ref{eq:piRho1loop}), which grow exponentially in (proper) time. This behavior can be verified explicitly by performing a 'memory expansion' of Eq.~(\ref{eq:FEOM}), i.e. evaluating the memory integrals on the right hand side of Eq.~(\ref{eq:FEOM}) around the latest (proper) times of interest \cite{JB:PR,SS:PR}. We find that the dominant contribution originates from the statistical self-energy in Eq.~(\ref{eq:piF1loop}), whereas the contribution from the spectral self-energy in Eq.~(\ref{eq:piRho1loop}) is effectively $\beta=2$ in the above classification scheme and thus becomes important only at later times. The modified evolution equations in this regime then take the form
\begin{eqnarray}
\label{eq:nlFEOM}
iG_{0,ac}^{-1,\mu\gamma}[x,\nu; A]F_{\gamma\nu}^{cb}(x,y,\Nu)=\frac{\delta_\tau^2}{2}\Pi^{(F),\mu\gamma}_{\textbf{-o-},ab}(x,y,\Nu)g_{\gamma\nu}(y)\;, \nonumber \\
\end{eqnarray}
where $\delta_\tau$ is the extent in time for which the memory integrals are evaluated. To obtain Eq.~(\ref{eq:nlFEOM}), we performed a leading order Taylor expansion of the integrand in Eq.~(\ref{eq:FEOM}) around $\tau_z=\tau_y$ and made use of the equal time commutation relations in Eq.~(\ref{eq:commrel}) to estimate the spectral function. The one-loop integral $\Pi^{(F)}$, which appears on the right hand side of Eq.~(\ref{eq:nlFEOM}), is dominated by the contributions from unstable modes and is proportional to $\exp[2\Gamma_0\sqrt{g^2\mu\tau}]$ at equal times. The contribution on the right hand side acts as a source term in the evolution equation of statistical fluctuations. As is well known from various examples of self-interacting quantum field theories, this term leads to a non-linear amplification of instabilities, where 'secondary' instabilities with strongly enhanced growth rates emerge over a large range of momenta. This has been observed in scalar field-theories \cite{JB:PR,SS:PR} as well as in non-abelian gauge theories \cite{JB:SU2,JB:SU3} and is a rather generic feature of self-interacting theories undergoing an instability. We will show in Sec.~\ref{sec:res} that the phenomenon of non-linear amplification also emerges in numerical simulations of the unstable Glasma and plays a crucial role in understanding gauge-invariant observables. The characteristic time scale for non-linear amplification to take place can be infered by comparing the magnitude of the non-linear contributions in Eq.~(\ref{eq:nlFEOM}) to the contributions of the background field. In the weak coupling limit this time scale is parametrically given by
\begin{eqnarray}
\label{eq:tSecParam}
\sqrt{g^2\mu\tau_{\text{Sec}}}\stackrel{g\ll1}{\sim}~\frac{1}{2\Gamma_0}\text{ln}~g^{-2} \;,
\end{eqnarray}
where $\Gamma_0$ is the characteristic growth rate of primary instabilities and we assumed $F_{\mu\nu}^{ab}\sim\mathcal{O}(1)$ at $\tau=0^{+}$ for characteristic modes. In addition to Eq.~(\ref{eq:tSecParam}) there are sub-leading contributions associated to the delayed set in of primary instabilities and the spectral distribution of statistical fluctuations in the initial state. The prior give rise to a constant contribution $\sqrt{g^2\mu\tau_{\text{Primary}}}$, while the latter enter only logarithmically in this estimate.
\\
\\
The emergence of secondary instabilities again modifies the power-counting, and one has to take into account also the contributions which originate from modes which exhibit secondary instabilities. Also higher order self-energy corrections become increasingly important as time proceeds and non-linear amplification can repeat itself, until at some point the growth of instabilities saturates and occupancies become as large as $\mathcal{O}(1/g^2)$. In this regime every truncation at a fixed loop order breaks down and the problem has to be addressed in a fully non-perturbative way. While in scalar quantum field theories there are different ways to address this problem, involving e.g. large $N$ resummation techniques \cite{JB:2PI1/N}, the most frequently employed approach in gauge theories is the classical-statistical approximation.
\section{Classical-statistical approximation}
\label{sec:classStat}
In the classical statistical approximation all fluctuations evolve classically and one neglects genuine quantum fluctuations, such as the ($\rho\rho$) quantum term in Eq.~(\ref{eq:piF1loop}). When transforming to the language of expectation values, as employed in the previous section, it can be shown that the prescription resums an infinite subset of diagrams \cite{JB:QvsCS}, such that the dynamics of fluctuations is treated on equal footing with the background fields. The set of diagrams included in the classical-statistical treatment can be identified as the self-energy corrections, which contain the most powers of the statistical propagator as compared to powers of the spectral function for each topology \cite{JB:QvsCS}. Accordingly, this corresponds to resumming the leading effects of the instability to all orders in the coupling constant \cite{CGC:IC}. In contrast to expansions at fixed loop orders, the classical statistical approximation thus provides a robust approximation scheme, which is particularly well suited for problems involving large statistical fluctuations. However there are problems associated with the Rayleigh-Jeans divergence, which concern the handling of ultra-violet divergences and the approach to thermal equilibrium at late times, which are discussed in more detail in the literature \cite{JB:QvsCS}.\\
\\
In the classical-statistical theory observables $\langle O(x)\rangle_{\text{cs}}$ are calculated as an ensemble average of classical field solutions $A_{\text{cl}}[A_{\tau_0},E_{\tau_0}]$, which individually satisfy the Yang-Mills evolution equations. The canonical field variables $A_{\tau_0}$ and $E_{\tau_0}$ at initial time $\tau_0$ are distributed according to a phase space density functional $W[A_{\tau_0},E_{\tau_0}]$, such that \cite{JB:QvsCS}
\begin{eqnarray}
\label{eq:expCS}
\langle O(x)\rangle_{\text{cs}}=&&\int DA_{\tau_0} DE_{\tau_0}~W[A_{\tau_0},E_{\tau_0}]~O_{cl}[A_{\tau_0},E_{\tau_0}]\;. \nonumber \\
\end{eqnarray}
Here $O_{\text{cl}}[A_{\tau_0},E_{\tau_0}]$ denotes the fact that the observable is evaluated as a functional of the classical field solution
\begin{eqnarray}
\label{eq:expClass}
O_{\text{cl}}[A_{\tau_0},E_{\tau_0}]=\int DA~O[A]~\delta(A-A_{\text{cl}}[A_{\tau_0},E_{\tau_0}]) \;, \nonumber \\
\end{eqnarray}  
where $A_{\text{cl}}[A_{\tau_0},E_{\tau_0}]$ is the classical Yang-Mills field solution 
with initial conditions $A_{\text{cl}}=A_{\tau_0}$ and $E_{\text{cl}}=E_{\tau_0}$ at initial time $\tau_0$. In practice, Eqns.~(\ref{eq:expCS}) and (\ref{eq:expClass}) state that vacuum fluctuations of the initial state are added on top of the background field at initial time, while the subsequent classical evolution keeps track of all non-linearities. The set of equations (\ref{eq:expCS}) and (\ref{eq:expClass}) is precisely the same as in Ref.~\cite{CGC:IC}, where it has been obtained as a partial resummation scheme of the perturbative corrections due to vacuum fluctuations of the initial state. The procedure outlined in Ref.~\cite{CGC:IC} consist of a hybrid approach, which employs the linearized evolution equations (\ref{eq:linA},~\ref{eq:linRho},~\ref{eq:linF}) outside the forward light-cone, where fluctuations are small, while switching to a classical-statistical description in the forward light-cone. This has the advantage that the evolution of the background field and the spectrum of fluctuations, which enter the phase-space weight $W[A_{\tau_0},E_{\tau_0}]$, can be obtained analytically. The phase-space average in Eq.~(\ref{eq:expCS}) can then be taken on the Cauchy surface $\tau_0=0^+$, such that only the dynamics in the forward light-cone has to be studied within classical-statistical lattice simulations. We follow this approach but employ a simpler spectrum of initial fluctuations, as specified in Sec.~\ref{sec:ic}, instead. This simplification is justified at sufficiently weak coupling, where the spectrum of fluctuations is quickly dominated by the growth of primary instabilities rather than the initial spectrum.
\subsection{Coupling dependence}
\label{sec:couplingDep}
An important property of the classical-statistical description is the independence of the gauge coupling constant $g$, in the sense that the classical-evolution equations are invariant under a change of variables
\begin{eqnarray}
\label{eq:rescaledF}
A_{\mu}^a(x)\rightarrow g~A_{\mu}^a(x)\;,\quad
F_{\mu\nu}^{ab}(x,y)\rightarrow g^2F_{\mu\nu}^{ab}(x,y)\;,
\end{eqnarray}
whereas the spectral function remains unaffected, as encoded in the equal time commutation relations.\footnote{In the classical theory $-i$ times the commutator is to be replaced by the Poisson brackets \cite{JB:QvsCS}.} With the rescaling (\ref{eq:rescaledF}) the entire coupling dependence in the classical-statistical evolution can be absorbed into the initial conditions at $\tau=0^+$, while the classical evolution equations become independent of the coupling constant. For the Glasma background fields this can be achieved most efficiently by replacing $gA_{\mu}^a(x)\rightarrow \tilde{A}_{\mu}^a(x)$ and $g\varrho^{(1/2)}(x_{\bot})\rightarrow\tilde{\varrho}^{(1/2)}(x_{\bot})$ for all expressions in Sec.~\ref{sec:classSol}, where in the MV model prescription
\begin{eqnarray}
\label{eq:MVmodelCoupl}
\langle \tilde{\varrho}^{(A)}_a(x_{\bot})\tilde{\varrho}{(B)}_b(y_{\bot})\rangle=g^4\mu^2\delta(x_{\bot}-y_{\bot})\;.
\end{eqnarray}
Here the model parameter $g^2\mu$ is directly related to the saturation scale $Q_s$ without further powers of the coupling constant appearing in the expression \cite{MVModel}. Accordingly the defining equation (\ref{eq:MVmodelCoupl}) is indeed independent of the value of the coupling constant. In contrast, the coupling constant $g$ appears explicitly in the initial spectrum of fluctuations, given by
\begin{eqnarray}
\left.\tilde{F}_{\mu\nu}^{ab}(x,y)\right|_{\tau=\tau'=0^+}=\left.g^2~F_{\mu\nu}^{ab}(x,y)\right|_{\tau=\tau'=0^+} 
\end{eqnarray}
and similarly for derivatives at initial time. Here it is important to note that the magnitude of the vacuum fluctuations $F_{\mu\nu}^{ab}(x,y)$ on the right hand side is independent of the value of the coupling constant. Thus the initial suppression of vacuum fluctuations compared to the boost invariant background fields is the only measure of the strong coupling constant, present in the classical-statistical field theory. We will exploit this fact in Sec.~\ref{sec:seed}, where we vary the amplitude of initial fluctuations in our simulations to study the coupling dependence of our results.
\begin{figure*}[t]
\includegraphics[width=0.7\textwidth]{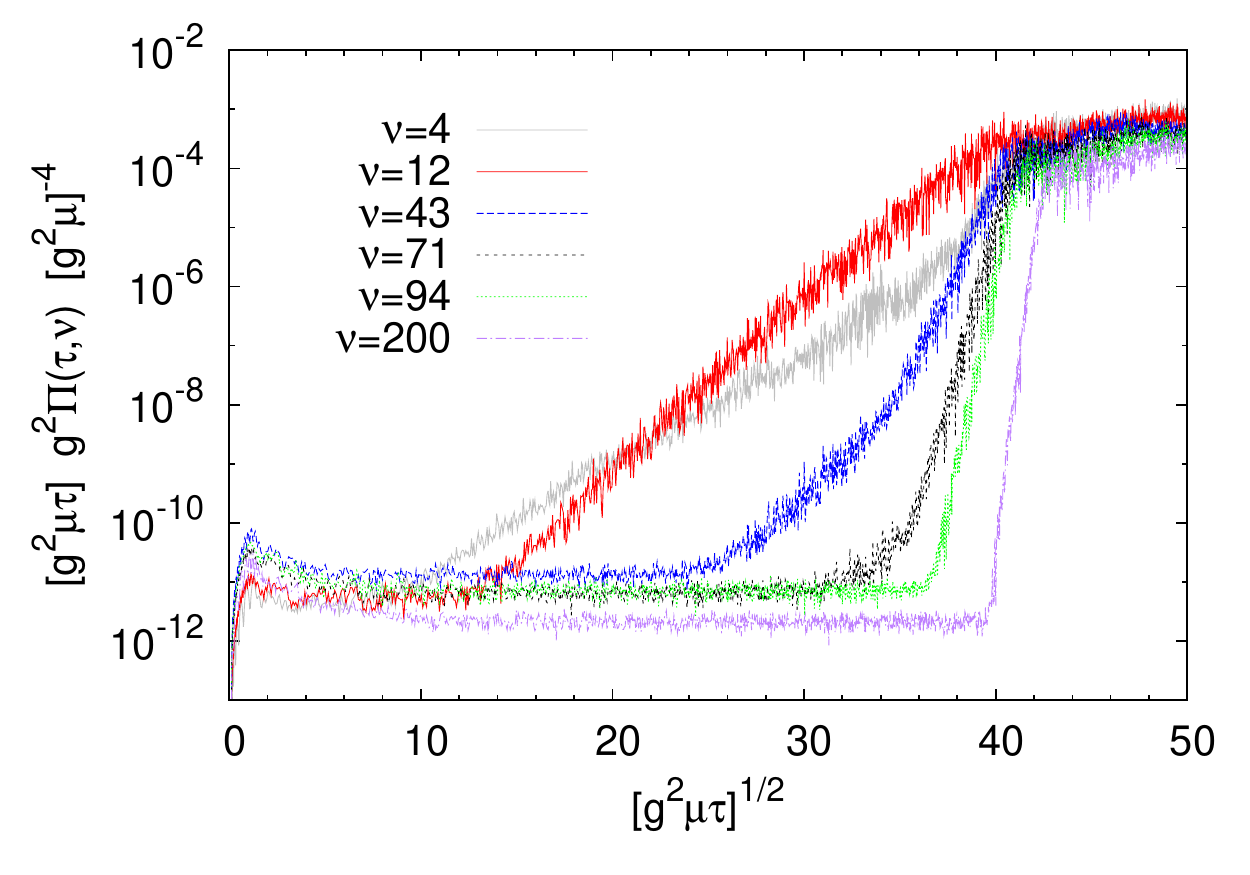}
\caption{ (color online) \label{fig:Modes} Time evolution of the pressure-pressure correlator $\Pi_L(\tau,\nu)$ for different rapidity wave numbers $\nu$. Once the initial fluctuations have grown larger one observes the emergence of the secondary instabilities, with growth rates. Subsequently the instability propagates towards higher momenta until saturation occurs and the system exhibits a much slower dynamics. The results are obtained for the MV model parameter $g^2\mu L_\bot=22.6$ on lattices with $N_\bot=16$ and $N_\eta=1024$ sites. The initial fluctuations are parametrized by $\Delta=10^{-10}$ and $b=0.01$ according to Eq.~(\ref{eq:initspec}).}
\end{figure*}
\subsection{Initial conditions at $\tau=0^+$}
\label{sec:ic}
Within the classical-statistical framework, the initial conditions for the time-evolution in the forward light-cone are given at $\tau=0^+$ by the set of equations (\ref{eq:IC1}) and (\ref{eq:alpha}), complemented by the spectrum of initial fluctuations. While in general it is necessary to implement the spectrum of fluctuations as specified in Ref.~\cite{CGC:IC}, it is also clear that at sufficiently weak coupling many details of the spectral shape of the initial fluctuations become irrelevant due to the presence of instabilities, which quickly dominate the spectrum. Since implementing the spectrum of fluctuations in Ref. \cite{CGC:IC} is numerically challenging, we will therefore stick to a simpler choice, where in accordance with previous works \cite{GlasmaRV,GlasmaGelis} the statistical fluctuations initially take the form
\begin{eqnarray}
\left.F_{\mu\nu}^{ab}(x,y)\right|_{\tau=\tau'=0}&=&0 \;,\\
\left.\partial_{\tau}F_{\mu\nu}^{ab}(x,y)\right|_{\tau=\tau'=0}&=&0 \;,\nonumber \\ 
\left.\partial_{\tau}\partial_{\tau'}F_{\mu\nu}^{ab}(x,y)\right|_{\tau=\tau'=0}&=&\langle\delta E_\mu^a(x_\bot,\eta_x)\delta E_\nu^b(y_\bot,\eta_y)\rangle \;.\nonumber
\end{eqnarray}
with
\begin{eqnarray}
\delta E_i^a(x_{\bot},\eta)&=&\partial_\eta f(\eta) e_i^a(x_{\bot}) \\
\delta E_\eta(x_{\bot},\eta)&=&-f(\eta) D_i e_i^a(x_{\bot})\;.
\end{eqnarray}
such that $\delta E_{\mu}^a(x)$ is an additive contribution to the background field $E_{\mu}^a(x)$ at initial time. The advantage of this construction is that the Gauss constraint is satisfied explicitly for arbitrary functions $f(\eta)$ and $e_i^a(x_{\bot})$. For the numerical simulations we choose
\begin{eqnarray}
\label{eq:initspec}
\langle e_i^a(p_T) e_j^b(q_T) \rangle&=&\Delta^2~\delta_{ij}\delta^{ab}\delta(p_T+q_T)\;, \\
\langle f(\nu)f(\nu')\rangle&=&e^{-2b|\nu|}\delta(\nu+\nu')\;,
\end{eqnarray}
where $p_T$ and $\nu$ are the Fourier coefficients with respect to  relative transverse coordinates and relative rapidity. Here $b$ is a (small) number, which regulates the ultra-violet divergence and the dimensionless parameter $\Delta$ controls the initial amplitude of small wave-number fluctuations. While this construction does not respect the details of the spectral composition, the parameter $\Delta$ provides a measure of the coupling constant $\Delta^2 \sim g^2$  (see Sec.~\ref{sec:couplingDep}) and we will vary its size in Sec.~\ref{sec:seed} to study the coupling dependence.
\section{Lattice Results}
\label{sec:res}
In this section we present results from classical-statistical lattice simulations of the Glasma evolution in the presence of boost non-invariant fluctuations.
 While the existence of a non-equilibrium instability has been established in previous simulations \cite{GlasmaRV,GlasmaGelis},
 we focus on the non-linear regime where unstable modes have grown large enough to significantly alter the dynamics. In contrast to the linear regime,
 where the initial size of boost non-invariant fluctuations is irrelevant for the dynamics of unstable modes, it is clear that for the non-linear regime the size of the initial fluctuations matters.
 In view of the spectrum of initial fluctuations obtained in Ref.~\cite{CGC:IC},
 the ratio of the initial amplitude of fluctuations compared to the amplitude of the (squared) background field is parametrically of the order of the strong coupling constant $g^2$.
 We study this dependence in Sec.~\ref{sec:seed} by considering different amplitudes of the initial fluctuations,
 as characterized by the dimensionless parameter $\Delta^2\sim g^2$ (see Sec.~\ref{sec:ic}).
 We restrict our analysis to weak coupling ($g^2\ll1$), where classical-statistical methods are expected to provide an accurate description of the quantum dynamics on large time scales.\\
\\
The discussion of our results is organized as follows: In Sec.~\ref{sec:instabilities} we investigate the dynamics of the instability at weak coupling and show how deviations from the linear regime emerge in terms of secondary instabilities. To further analyze this behavior, we obtain the relevant growth rates for primary and secondary instabilities as well as the corresponding set-in times. Subsequently, in Sec.~\ref{sec:seed}, we investigate the dependence on the coupling constant by varying the size of initial fluctuations.\\
\\
If not stated otherwise we perform simulations on $N_\bot=16$, $N_{\eta}=1024$ and $N_\bot=32$, $N_{\eta}=128$ lattices and we employ the set of parameters $g^2\mu~N_\bot a_\bot=22.6$ and $N_{\eta} a_{\eta}=1.6$ in accordance with Ref.~\cite{GlasmaRV}. We study the time evolution of the gauge-invariant pressure-pressure correlation function 
\begin{eqnarray}
\Pi_L^2(\tau,\eta,\eta')&=&\left< P_L(\tau,x_\bot,\eta)P_L(\tau,y_\bot,\eta')\right>_T\;,
\end{eqnarray}
where $P_L(x)$ is the longitudinal pressure as a function of space and time arguments and $<.,.>_T$ denotes average over transverse coordinates and classical-statistical ensemble average. We will frequently employ the Fourier transforms of $\Pi_L^2(\tau,\eta,\eta')$ with respect to relative rapidity, i.e. we consider
\begin{eqnarray}
\label{eq:obsdef}
\Pi_L^2(\tau,\nu)&=&(N_{\eta}a_{\eta})^{-1}\int d\eta~d\eta'~\Pi_L(\tau,\eta,\eta')~e^{-i\nu(\eta-\eta')}\;, \nonumber \\
\end{eqnarray}
and usually show results for $\Pi_L(\tau,\nu)$, i.e. the square root of the above expression. 
The details of our lattice setup are described in more detail in the appendix.
\subsection{The unstable Glasma}
\begin{figure}[t]
\includegraphics[width=0.48\textwidth]{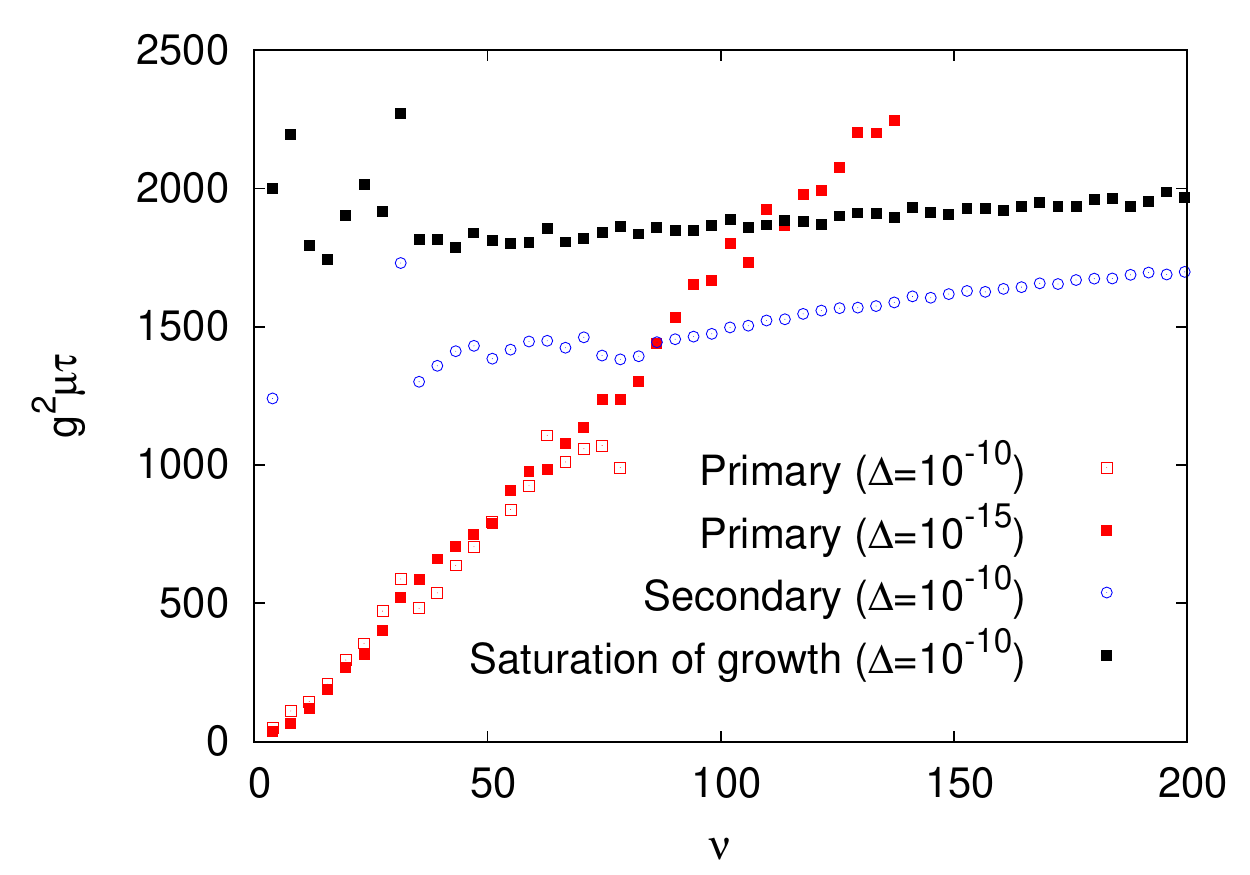}
\includegraphics[width=0.48\textwidth]{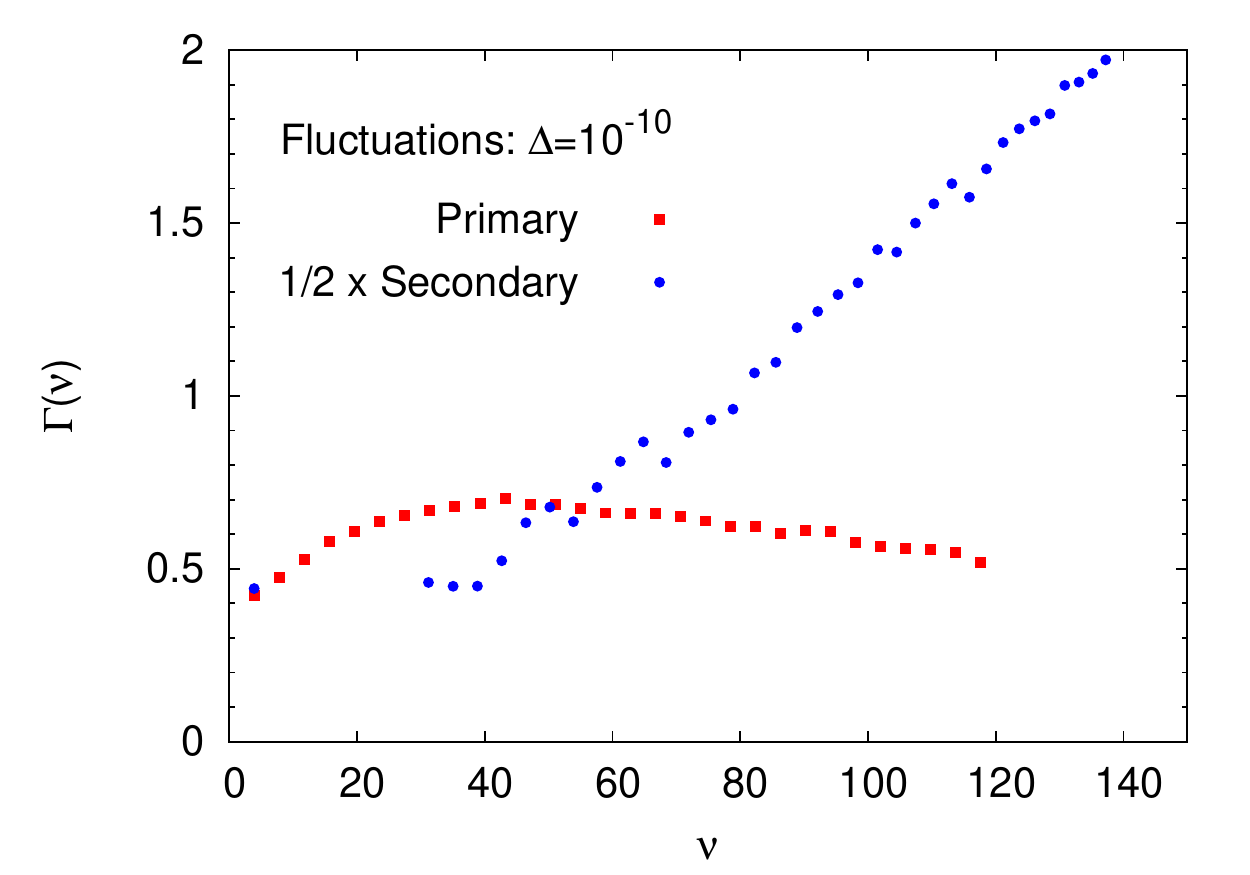}
\caption{ (color online) \label{fig:fits} (\textbf{top}) Set in times of primary and secondary instabilities as a function of rapidity wave number $\nu$. Once secondary instabilities set in, the growth quickly extends to higher rapidity wave numbers.  (\textbf{bottom}) Growth rates $\Gamma(\nu)$ as a function of rapidity wave number $\nu$ for primary and secondary instabilities. The results are obtained for the MV model parameter $g^2\mu L_\bot=22.6$ on lattices with $N_\bot=16$ and $N_\eta=1024$ sites.}
\end{figure}
\label{sec:instabilities}
In a first step we study the time evolution of the gauge-invariant pressure-pressure correlator $\Pi_L(\tau,\nu)$ for a fixed value of the amplitude of initial fluctuations $\Delta=10^{-10}$ and $b=0.01$. The results are shown in Fig.~\ref{fig:Modes} for different rapidity wave numbers $\nu$ as a function of time. From Fig.~\ref{fig:Modes} one observes a sequence of different dynamical regimes which are characterized as follows:\\
\\
At very early times $\sqrt{g^2\mu\tau}\lesssim2$ one observes a period of rapid initial growth, which is presumably caused by the dephasing dynamics of the strong background fields, that takes place roughly on the same time scale \cite{GlasmaGelis,CGC2DKrasnitz,CGC2DMcLerran,CGC2DBlaizot,CGC2DLappi,ToyGlasmaFuji}. However at weak coupling, i.e. for small fluctuations, this constitutes a rather small effect as the unstable modes exhibit their dominant growth at later times.\\
\\
The rapid initial period is followed by a regime where the Glasma instability \cite{GlasmaRV,GlasmaGelis} is operative and modes with non-zero rapidity wave number exhibit exponential amplification. The instability sets in with a delay for higher momentum modes and the functional form is well described by an exponential of the form $\exp[\Gamma(\nu)\sqrt{g^2\mu\tau}]$, with the momentum dependent growth rate $\Gamma(\nu)$, as seen for $\nu=4,12$ in Fig.~\ref{fig:Modes}. To further investigate this behavior we fit a set of continuous piecewise linear functions to the modes displayed in Fig.~\ref{fig:Modes} in order to obtain the relevant growth rates and set-in times. The results of these fits are shown in Fig.~\ref{fig:fits} as a function of rapidity wave number $\nu$. From the upper panel of Fig.~\ref{fig:fits} one observes that the primary set-in times follow a linear behavior, as reported in Ref.~\cite{GlasmaRV}. The primary growth rates are shown in the lower panel of Fig.~\ref{fig:fits}. One observes that modes with small rapidity wave number exhibit smaller growth rates as compared to modes with higher rapidity wave number, while at large $\nu$ the primary growth rates become approximately constant. The numerical values are compatible with the results reported in Ref.~\cite{GlasmaRV}, where characteristic growth rates were obtained from a convolution of the spectrum.\\
\\
While the primary instability continues to set-in for higher momentum modes, one observes from Fig.~\ref{fig:Modes} that at later times modes with intermediate ($\nu=43,~71$) and small ($\nu=4$) rapidity wave number suddenly exhibit much higher growth rates than previously observed. This change in the dynamics becomes evident when shortly after modes with even higher rapidity wave numbers ($\nu=94,~200$) exhibit even stronger growth rates, such that the spectrum extends quickly to the ultra-violet and the instability propagates towards higher momenta. This is precisely the signature of secondary instabilities, where non-linear self-interactions among unstable modes give rise to an amplification of the primary instability. The amplification happens initially in a small momentum region and then quickly propagates outwards to higher momenta. This can be seen in the upper panel of Fig.~\ref{fig:fits}, where we show the set-in times of primary and secondary growth. One also observes that for modes with large rapidity wave number $\nu>\nu_{\text{c}}$ secondary instabilities set-in before the primary instability, such that the growth of high $\nu$ modes is solely due to non-linear effects. The numerical value of $\nu_{\text{c}}$ depends, of course, on the size of the initial fluctuations and we will confirm the non-linear origin of this phenomenon in Sec.~\ref{sec:seed}, where we investigate in more detail the dependence on the initial amplitude of fluctuations. The dynamical power counting scheme developed in Sec.~\ref{sec:pc} suggests that these secondary instabilities are caused by the one-loop diagram shown in the upper panel of Tab.~\ref{tab:pc}, with secondary growth rates as large as twice the primary ones. If we compare the rates of primary and secondary growth, as shown in the lower panel of Fig.~\ref{fig:fits}, we find that this is indeed the case for modes with intermediate $\nu$, which exhibit the earliest non-linear amplification. Modes with higher values of $\nu$ exhibit even larger growth rates, which can be attributed to multiple amplification processes as well as higher order corrections.\\
\\
The growth of primary and secondary instabilities in Fig.~\ref{fig:Modes} continues until at some point saturation of the instability sets in and the system has reached non-perturbatively large occupation numbers. In this regime we observe that the process of non-linear amplification continues even after the growth of the leading primary modes has saturated. This has a significant impact also on bulk observables such as the ratio of longitudinal pressure to energy density. Before we turn to a more detailed discussion of this highly occupied regime, we will first investigate the coupling dependence of the non-linear amplification process.

\subsection{Coupling dependence}
\label{sec:seed}
In Sec.~\ref{sec:instabilities} we have discussed the time evolution of non-boost invariant fluctuations in the Glasma. We have shown that, at weak coupling, primary instabilities of small rapidity wave number fluctuations occur. In turn, these cause secondary instabilities of modes with higher rapidity wave numbers until saturation of the instabilities occurs and one enters the highly over-occupied regime. In this section we investigate in more detail the dependence on the choice of parameters, in particular the impact of the initial amplitude of fluctuations. According to Sec.~\ref{sec:couplingDep}, this can be interpreted as varying the value of the coupling constant $g^2$ in our simulations, without respecting in detail the spectral shape of initial fluctuations \cite{CGC:IC}.\\
\\
\begin{figure}[t]
\includegraphics[width=0.48\textwidth]{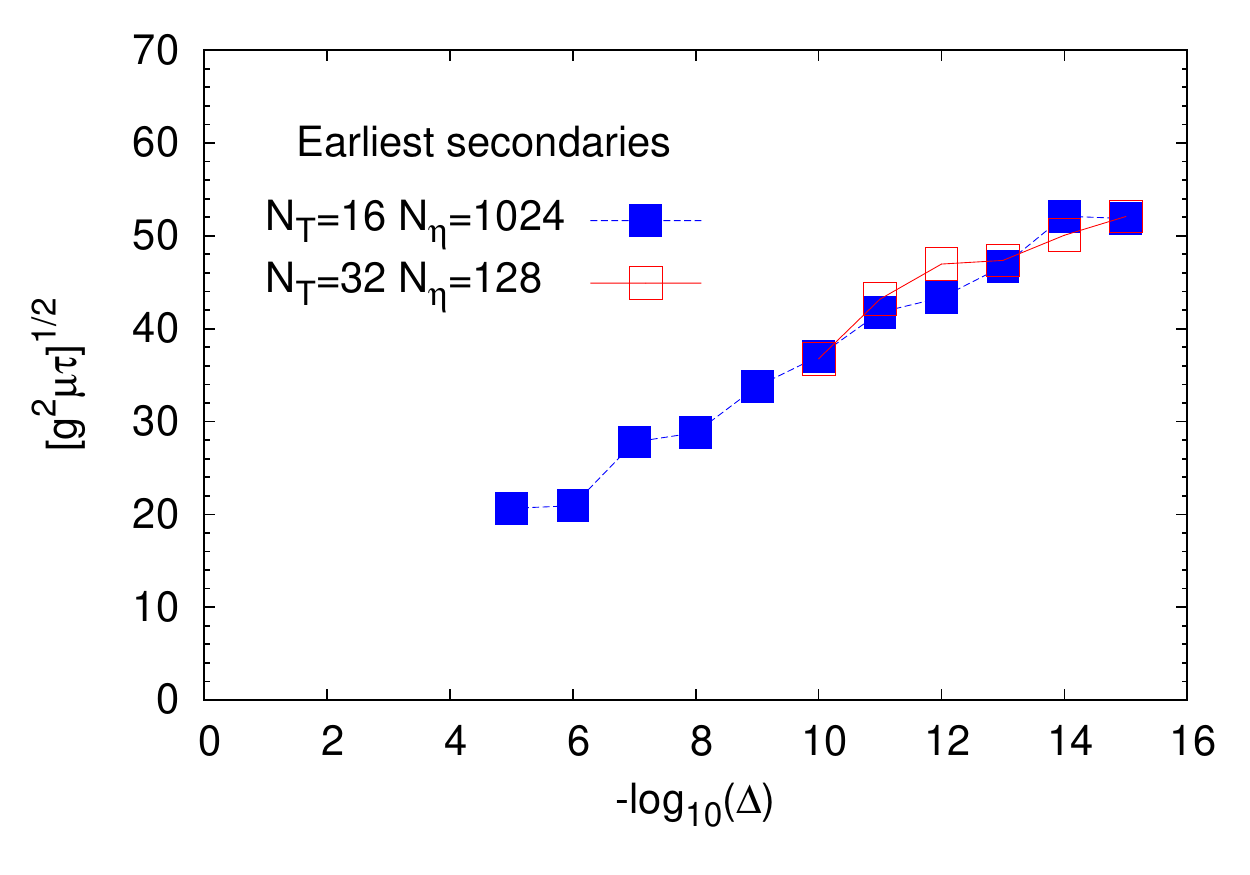}
\caption{ (color online) \label{fig:Delta} 
Set in times of secondary instabilities as a function of the size of initial fluctuations. The parameter $\Delta$ quantifies the magnitude of initial fluctuations and is proportional to the coupling constant $g$ (see also Sec.~\ref{sec:ic}). One observes a logarithmic dependence in accordance with analytic expectations. The results are obtained for the MV model parameter $g^2\mu L_\bot=22.6$.}
\end{figure}
In order to study the dependence on the initial amplitude of fluctuations, we vary the parameter $\Delta$ in the range of $10^{-15}$ to $10^{-5}$. The qualitative behavior is the same as observed in Sec.~\ref{sec:instabilities} for all values of $\Delta$, i.e. we observe primary instabilities followed by non-linear amplification and subsequent saturation of the growth. Due to the non-linear origin, the time scales for the set in of secondary instabilities and the saturation of growth depend, of course, on the initial amplitude of fluctuations. The results of our analysis are shown in Fig.~\ref{fig:Delta}, where we show the characteristic set-in times of secondary instabilities as a function of the initial amplitude $\Delta$ of boost non-invariant fluctuations. The rather weak dependence observed in Fig.~\ref{fig:Delta} stems from the fact that, at early times, the magnitude of non-linear contributions depends exponentially as $\exp[2\Gamma_0\sqrt{g^2\mu\tau}]$, whereas the dependence on the initial amplitude is just a power. Assuming that non-linear amplification is caused by the one-loop diagram depicted in Tab.~\ref{tab:pc} we obtain the parametric estimate (see Sec.~\ref{sec:pc})
\begin{eqnarray}
\sqrt{g^2\mu\tau_{\text{Secondary}}}\sim\sqrt{g^2\mu\tau_{\text{SetIn}}}+\frac{1}{2\Gamma_0}\text{ln}(g^{-2})\;, 
\end{eqnarray}
where $\tau_{\text{SetIn}}$ characterizes the set-in time of primary instabilities and $\Gamma_0$ is the characteristic primary growth rate. This behavior is reproduced by the lattice data on a qualitative level.
\subsection{Saturated regime}
\label{sec:BULK}
\begin{figure}[t]
\includegraphics[width=0.48\textwidth]{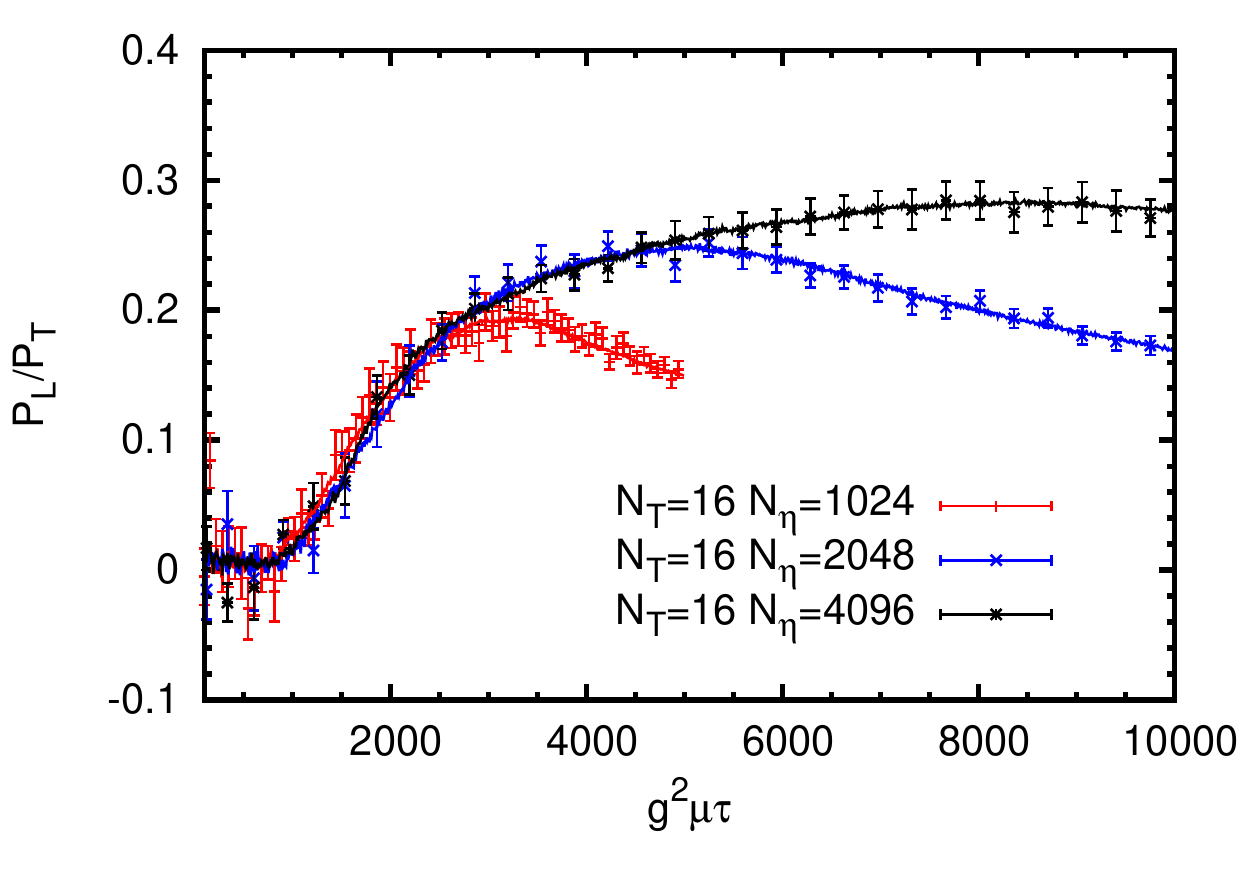}
\caption{ (color online) \label{fig:PLvsPT} Ratio of longitudinal and transverse pressure as a function of time. The different curves correspond to different values of the lattice spacing. One observes a remaining order one anisotropy over a large time scale. The results are obtained for the MV model parameter $g^2\mu L_\bot=22.6$ and the initial fluctuations are chosen as $\Delta=10^{-5}$ and $b=0.1$.}
\end{figure}

The evolution of the system in the saturated regime is of great interest, when studying the thermalization process at weak coupling. In this regime the system exhibits a much slower dynamics and one expects the system to become approximately isotropic on sufficiently large time scales \cite{BMSS,Moore,Blaizot}. In order to analyze the behavior, we first consider the evolution of the ratio of longitudinal pressure to energy density, as a measure of the bulk anisotropy of the system. The challenge in this analysis comes from the fact that the relevant ultra-violet cutoff associated with longitudinal momentum $\Lambda_z\sim\pi/(\tau a_\eta)$ decreases with (proper) time. Furthermore having a large rapdity cutoff $\Lambda_\eta\sim\pi/a_\eta$ can cause severe problems at early times and a proper renomalization scheme might be needed to ensure physical results. We adress this problem by chosing the initial amplitude of fluctuations very small, such that the overall contribution of fluctuations to the energy density is less than a percent even for the largest cutoffs that we consider. We then vary the lattice spacing $a_\eta$ while keeping $N_{\eta}a_{\eta}$ fixed to study the sensitivity to the cutoff. The results are presented in Fig.~\ref{fig:PLvsPT}, where we show the ratio of longitudinal and transverse pressure as a function of time. While at early times the longitudinal pressure of the system is consistent with zero, we observe a clear rise of the longitudinal pressure towards later times. In the saturated regime the trend towards isotropization slows down dramatically and the system exhibits a remaining order one anisotropy over a large time scale. The results are insensitive of the longitudinal discretization, as long as the lattice spacing $a_\eta$ is sufficiently small.\\
\begin{figure}[t]
\includegraphics[width=0.48\textwidth]{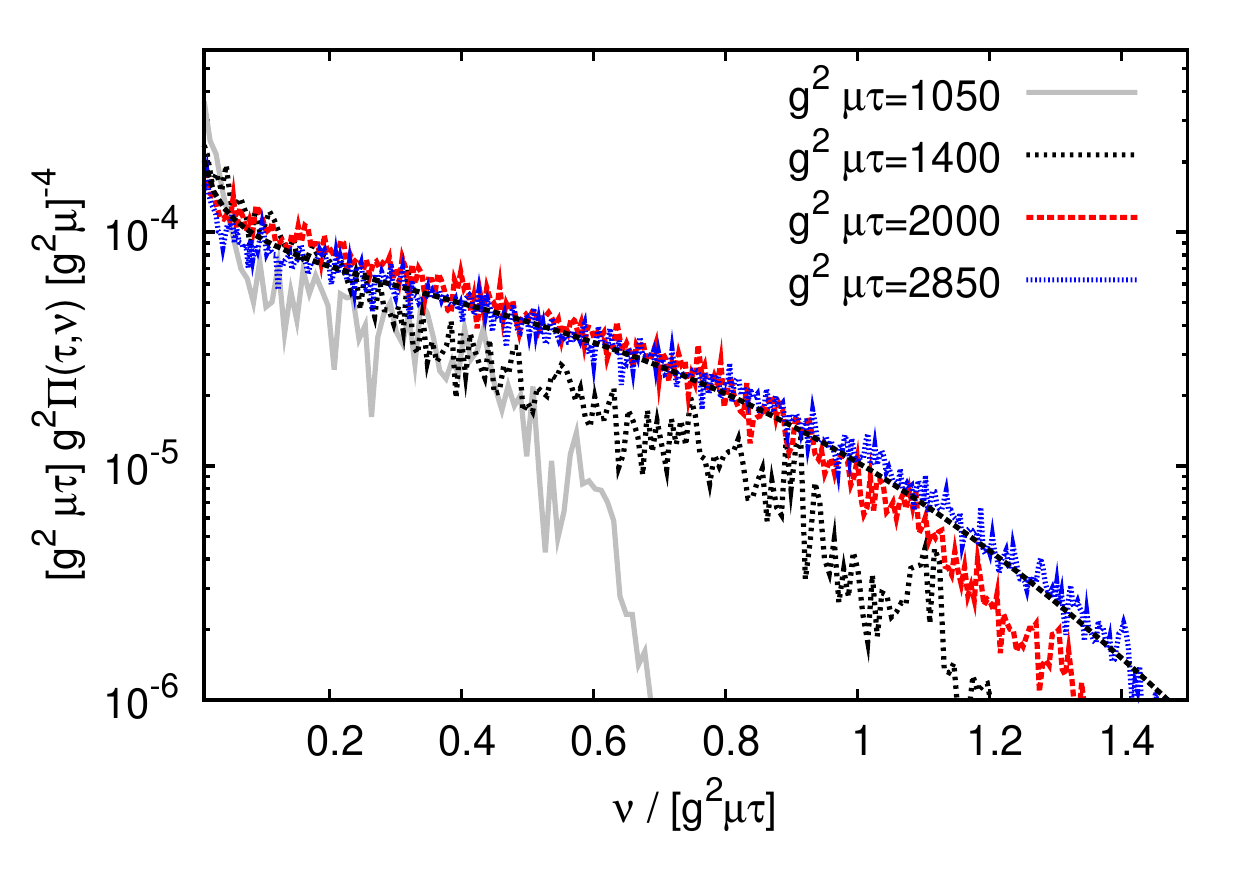}
\includegraphics[width=0.48\textwidth]{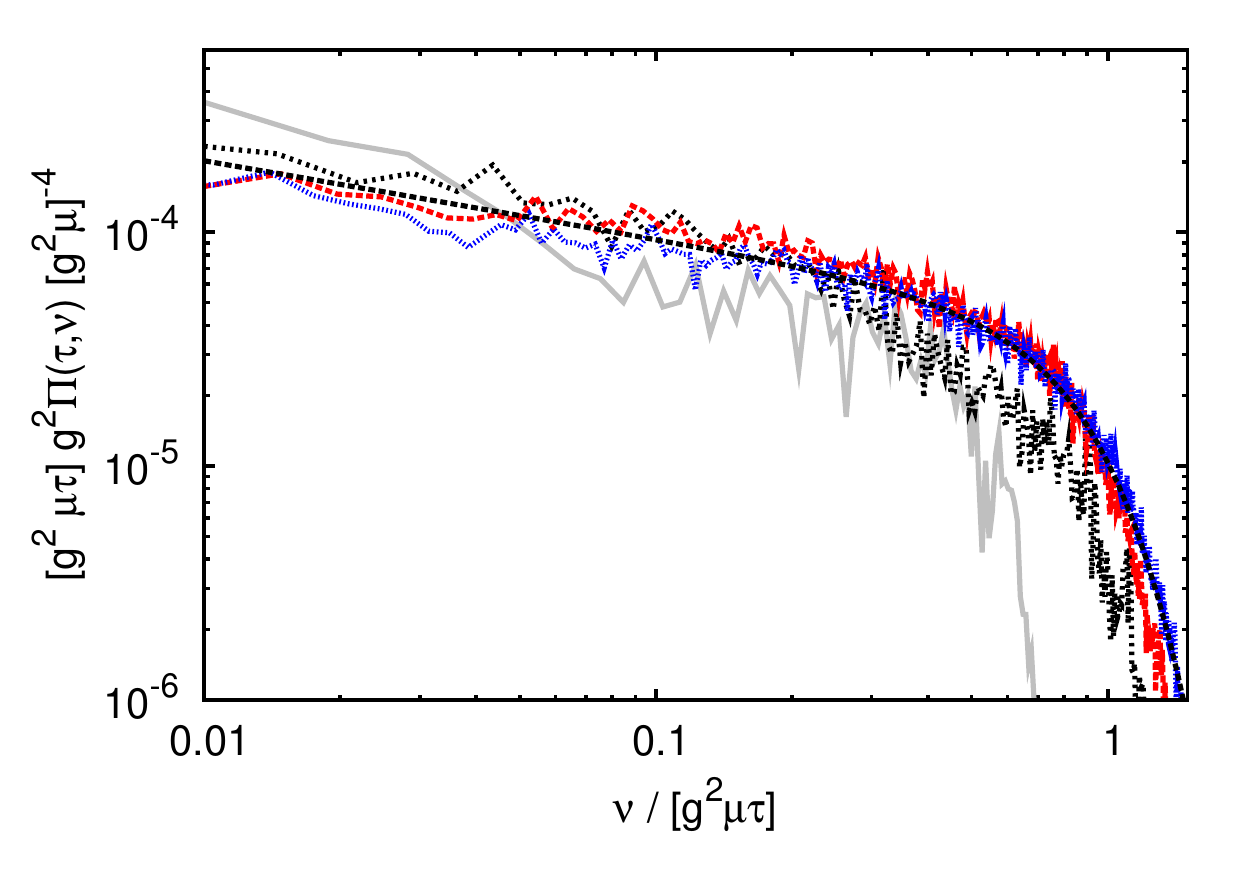}
\caption{ (color online) \label{fig:Spectra} Spectrum of the pressure-pressure correlation function $\Pi(\tau,\nu)$ as a function of longitudinal momenta $p_{z}=\nu/\tau$ at different times of the evolution. The top panel shows the spectrum on a log-linear plot, whereas the lower panel corresponds to a double logarithmic plot. The black dashed line corresponds to a fit of the functional form (\ref{eq:fit}).}
\end{figure}
\\
We also study the spectrum of the pressure-pressure correlation function $\Pi(\tau,\nu)$ in this regime. This is shown in Fig.~\ref{fig:Spectra} at different times, as a function of longitudinal momentum $p_z=\nu/\tau$. From the top panel of Fig.~\ref{fig:Spectra}, one observes how the spectrum rapidly extends towards higher momenta at times $g^2\mu\tau\simeq1000-2000$, while the amplitude at low momenta decreases. This redistribution is accompanied by the increase in the bulk pressure observed in Fig.~\ref{fig:PLvsPT}. At later times $g^2\mu\tau\gtrsim2000$ the evolution of the spectrum features enhanced contributions from soft modes and a strong fall-off at high momenta. The evolution of the spectrum in this regime proceeds in a much slower way. The lower panel of Fig.~\ref{fig:Spectra} shows the spectrum on a double logarithmic plot. One observes that the soft tail of the spectrum can be described by a power-law. This is illustrated by the black dashed line in Fig.~\ref{fig:Spectra}, which corresponds to the functional form
\begin{eqnarray}
\label{eq:fit} 
f(x)=A~x^{-\beta}/(1+\exp[B x^2])\;,
\end{eqnarray}
i.e. a power-law with normal UV regulator. This behavior is very similar to wave-turbulence observed in non-expanding systems \cite{JB:turb,SS:turb,Moore:turb}. The power law features an infrared power-law exponent of $\beta\simeq1/3$, while the regulator controls the fall-off at high momenta. The shape of the spectrum at later times is qualititive similar. However we already find sizable amplitudes for modes with longitudinal momenta on the order of the transverse lattice cutoff. To avoid discretization errors, one therefore has to consider also much larger lattices in the transverse direction and we expect future simulations to extent the studies of this regime. 

\section{Conclusion}
\label{sec:conclusion}
In this paper we discussed the impact of quantum fluctuations on the non-equilibrium dynamics of the Glasma. The picture that emerges at weak coupling is quite universal and depends only weakly on the details of initial fluctuations. In the weak coupling scenario initially boost non-invariant fluctuations exhibit exponential growth until at some point they have grown large enough for non-linear interactions to become important. At this stage secondary instabilities set in for a large momentum region, which extends to much higher rapidity wave numbers as the primary growth. Both primary and secondary growth continues until saturation of the instability occurs and the system exhibits a much slower dynamics. This scenario is similar to various examples of instabilities in self-interacting quantum field theories \cite{JB:PR,SS:PR,JB:SU2,JB:SU3} and we have shown in Sec.~\ref{sec:pc} how the emergence of secondary instabilities can be studied systematically within the framework of two particle irreducible effective action techniques. At the qualitative level this analysis is sufficient to predict the parametric dependence of the set-in time of secondary instabilities and the estimate can be made quantitative when the microscopic dynamics of the primary instability is described analytically. The set in time of secondary instabilities depends on the growth rates of the primary instability and to a much weaker extent on the size of the initial fluctuations, which is related to the value of the strong coupling constant. We confirmed this qualitative behavior by varying the size of the initial fluctuations in the classical-statistical lattice simulations, without respecting in detail the spectral composition. The latter will be taken into account in future studies, though one does not expect qualitative changes at weak coupling, where the longitudinal spectrum is quickly dominated by the exponential growth of primary instabilities. In contrast, changes in the spectrum of the background field may have a much more significant impact on the dynamics, as they readily alter the dynamics of primary instabilities. In particular this may change the character of the instability from the Weibel type, which is present in the MV model, to the Nielsen Olesen type \cite{ToyGlasmaFuji,SS:NO} and it will be important to consider more realistic saturation models in the future. This is important also in view of the applicability at RHIC and LHC energies, where in addition one has to consider much larger values of the strong coupling constant. This is not unambiguous since one encounters conceptual problems concerning the renormalization of the theory as well as the impact of sub-leading quantum corrections. While exploratory studies in scalar field theories have recently obtained promising results \cite{ExpScalarsGelis,ExpScalarsHatta}, we expect future studies of non-abelian gauge theories to expand on this issue.\\
\\
In addition to the unstable regime, we also studied the dynamics of the saturated regime, which is of great interest in the recent debate on the thermalization mechanism at weak coupling and high collider energies \cite{BMSS,Moore,Blaizot}. In this context, different scenarios involving elastic and inelastic scattering proccesses \cite{BMSS}, instability induced isotropization \cite{Moore} as well as the formation of a transient condensate \cite{Blaizot} have been proposed and are currently under investigation. While for non-expanding systems, the results from classical-statistical lattice simulations suggest the occurrence of a turbulent cascade \cite{JB:turb,SS:turb,Moore:turb}, there are only few results for expanding systems \cite{GlasmaGelis,Strickland} and we expect more studies in the future.  While a dedicated lattice study has the potential to clarify these questions, we restricted our analysis to the characteristic properties of the system at later times. We investigated the ratio of longitudinal and transverse pressure as a measure of the bulk anisotropy of the system. This observable can be used to distinguish between different scenarios which predict a characteristic evolution of this quantity. Our results indicate that the system exhibits an order one anisotropy on large time scales, which is common to the early stages of all the scenarios \cite{BMSS,Moore,Blaizot}. In order to clearly indentify the onset of an attractor solutions one therefore has to investigate even larger time scales and carefully monitor all discretization errors, which will be addressed in future studies. By investigating the spectrum of the pressure-pressure correlation function, we found evidence for a scaling solution and similar results have also been obtained in Ref.~\cite{GlasmaGelis}, where a different correlation function has been considered. The question whether this behavior is indeed related to the emergence of a turbulent cascade as observed in non-expanding systems \cite{JB:turb,SS:turb,Moore:turb}, will also be subject to future studies.\\
\\
\textit{Acknowledgment:} We thank F. Gelis, Y. Hatta, A. Kurkela, G.D. Moore, M. Strickland and R. Venugopalan for discussions. This work was supported in part by the BMBF grant 06DA9018.
\section*{Appendix: Classical-statistical lattice simulations}
\label{sec:lattice}
The classical-statistical lattice simulations are performed in the Hamiltonian framework with the lattice link variables $U_{\mu}(x)$ which are related to the continuum variables by
\begin{eqnarray}
\label{eq:linkvar}
U_{i}(x)&=&\exp[iga_{\bot}~t^aA_{i}^a(x)] \;, \\ U_{\eta}(x)&=&\exp[iga_\eta~t^a A_{\eta}^a(x)] \;,
\end{eqnarray}
where $a_{\bot}$ and $a_{\eta}$ are the lattice spacings in the transverse and longutidinal directions and $t^a$ denote the generators of the $SU(2)$ gauge group. The $SU(2)$ exponential can be computed as
\begin{eqnarray}
\exp[it^a A^a]=\cos(\sqrt{A^2}/2)\dblone+\frac{\sin(\sqrt{A^2}/2)}{\sqrt{A^2}}A^a t^a \;.
\end{eqnarray}
In Fock-Schwinger gauge one finds $U_{\tau}=\dblone$ and the (dimensionless) electric field variables read
\begin{eqnarray}
E_{i}^a(x)&=&ga~\tau\partial_{\tau}A_i^a(x)\;, \\
E_{\eta}^a(x)&=&ga^2~\tau^{-1}\partial_{\tau}A_\eta^a(x)\;.
\end{eqnarray}
Without loss of generality we set the coupling constant $g=1$ in the following  (see Sec.~\ref{sec:couplingDep}). The (dimensionless) lattice Hamiltonian density $\mathcal{H}(\tau,x_{\bot},\eta)$ of the system is given by
\begin{eqnarray}
\tau^{-1}\mathcal{H}=c_E\frac{\mathcal{E}_T^2}{2}+\frac{\mathcal{E}_L^2}{2}+\frac{\mathcal{B}_L^2}{2}+c_B\frac{\mathcal{B}_T^2}{2}
\end{eqnarray}
where $c_E=a^2/\tau^2$ and $c_B=a^2/(\tau^2 a_\eta^2)$, and the squares of the electric and magnetic field strengths are, for the $SU(2)$ gauge-group, given by
\begin{eqnarray}
\label{eq:EandB}
\mathcal{E}_T^2(x)&=&[E_i^a(x)]^2\;, \qquad \mathcal{E}_L^2(x)=[E_\eta^a(x)]^2 \\
\mathcal{B}_T(x)&=&4~\text{tr}\left[\dblone-U^{\Box}_{i\eta}(x)\right] \;, \quad
\mathcal{B}_L(x)=4~\text{tr}\left[\dblone-U^{\Box}_{12}(x)\right]\;. \nonumber
\end{eqnarray}
where $U_{\alpha\beta}(x)$ are the standard plaquettes given by
\begin{eqnarray}
U^{\Box}_{\alpha\beta}(x)&=&U_{\alpha}(x)U_{\beta}(x+\hat{\alpha})U^{\dagger}_{\alpha}(x+\hat{\beta})U_{\beta}^{\dagger}(x) \\
V^{\Box}_{\alpha\beta}(x)&=&U_{\alpha}(x)U^{\dagger}_{\beta}(x+\hat{\alpha}-\hat{\beta})U^{\dagger}_{\alpha}(x-\hat{\beta})U_{\beta}(x-\hat{\beta}) \nonumber
\end{eqnarray}
where $\alpha,\beta=1,2,\eta$ and we also defined the $V^{\Box}_{\alpha\beta}$ plaquettes for later use.

\subsection{Evolution equations}
The Hamiltonian evolution equations on the lattice in co-moving ($\tau,\eta$) coordinates are given by the equation of motion for the link variables
\begin{eqnarray}
\label{eq:UP1}
U_{\alpha}(x_T,\eta,\tau+a_\tau)=W^{\Box}_{\alpha}(x_T,\eta,\tau+a_\tau/2)U_{\alpha}(x_T,\eta,\tau)\;, \nonumber \\
\end{eqnarray}
(no summation over $\alpha=1,2,\eta$). Here $W^{\Box}_{\alpha}$ denote plaquettes involving a time link, given by
\begin{eqnarray}
W^{\Box}_{\alpha}(x)=\exp\left[it^a~c_{\alpha}E_{\alpha}^a(x)\right]\;,
\end{eqnarray}
(no summation over $\alpha=1,2,\eta$) and the coefficients read
\begin{eqnarray}
c_i=\frac{a_{\tau}}{\tau} \;, \quad c_\eta=\frac{\tau a_\eta a_\tau}{a^2}\;.
\end{eqnarray}
The update rules for the chromo-electric fields are given by
\begin{eqnarray}
\label{eq:UP2}
E_{\alpha}^a(x_T,\eta,\tau+a_\tau/2)&=&E_{\alpha}^a(x_T,\eta,\tau-a_\tau/2) \\
&+&2~d_{\alpha\beta}~\text{tr}\left[it^a \left(U^{\Box}_{\alpha\beta}(x)+V^{\Box}_{\alpha\beta}\right)(x)\right] \nonumber
\end{eqnarray}
with the coefficients
\begin{eqnarray}
d_{ij}=\frac{a_\tau \tau}{a^2}\;, \quad d_{i\eta}=\frac{a_\tau}{\tau a_\eta^2} \;,
\quad d_{\eta i}=\frac{a_\tau}{\tau a_\eta} \;.
\end{eqnarray}
The time evolution can then be computed by alternately solving Eqns.~(\ref{eq:UP1}) and (\ref{eq:UP2}). The Gauss law constraint is conserved by this evolution and reads
\begin{eqnarray}
\frac{a}{a_\tau}\sum_{\alpha=1,2,\eta} h_{\alpha} \mathcal{G}^a_{\alpha}(x)&=&0\;,
\end{eqnarray}
which is satisfied separately for all $x$ and $a$. Here we denote $h_i=1$, $hd_{\eta}=a^2/(\tau^2a_\eta^2)$ and
\begin{eqnarray}
\mathcal{G}_{\alpha}^a(x)&=&\text{tr}\left[it^a~W^{\Box}_\alpha(x)\right] \\
&-&\text{tr}\left[it^a~U_\alpha^{\dagger}(x-\hat{\alpha})W^{\Box}_\alpha(x-\hat{\alpha})U_\alpha(x-\hat{\alpha})\right]\;. \nonumber
\end{eqnarray}

\subsection{Initial conditions on the lattice}
We first generate sets of uncorrelated standard Gaussian random numbers for every position in the transverse plane and every color associated to the color-charge densities of the nuclei, i.e.
\begin{eqnarray}
\tilde{\rho}_{a}^{(A)}(x_{\bot})=g^2\mu a~(\xi_{a}^{(A)}(x_{\bot})-R_a^{(A)})
\end{eqnarray}
where $\xi_{a}^{(A)}(x_t)$ are Gaussian random number and the subtraction of the overall color charge $R_a^{(A)}=N_\bot^{-2}\sum_{x_{\bot}}\xi_a^{(A)}(x_{\bot})$ ensures the overall color neutrality constraint. The result is then Fourier transformed to momentum space where we solve the Laplace equation
\begin{eqnarray}
\Lambda_a^{(A)}(p_T)=-p_T^{-2}\tilde{\rho}_a^{(A)}(p_T)\;.
\end{eqnarray}
The result is Fourier transformed back to obtain the solution of the Laplace equation in coordinate space. We then proceed by calculating the pure gauge solutions $U^{(1/2)}(x_\bot)$ according to
\begin{eqnarray}
U^{(A)}_{i}(x_\bot)&=&V^{(A)}(x_\bot)V^{\dagger(A)}(x_\bot+\hat{\imath})\;,\\
V^{(A)}(x_\bot)&=&\exp[it^a\Lambda_a^{(A)}(x_\bot)]\;.
\end{eqnarray}
Finally the link variables $U_{\mu}(x_\bot,\eta)$  at initial time are obtained as \cite{GlasmaRV,GlasmaGelis}
\begin{eqnarray}
U_{i}(x_\bot,\eta)&=&M_{i}(x_\bot)N_{i}(x_\bot)\;,\quad U_{\eta}(x_\bot,\eta)=\dblone\;,\\
M_{i}(x_\bot)&=&\left[U^{(1)}_i(x_\bot)+U^{(2)}_i(x_\bot)\right]\;, \\
N_{i}(x_\bot)&=&\left[U^{\dagger(1)}_i(x_\bot)+U^{\dagger(2)}_i(x_\bot)\right]^{-1}\;. 
\end{eqnarray}
and the electric fields $E_{\mu}^a(x_\bot,\eta)$ are given by \cite{GlasmaRV}
\begin{eqnarray}
E_{\eta}^a(x_\bot,\eta)&=&\text{tr}[it^a \sum_{i=1,2} U^{(2)}_{i}(x_\bot)-U^{(2)}_{i}(x_\bot-\hat{\imath})  \\
&+&U^{(1)}_{i}(x_\bot)U^{\dagger}_i(x\bot)-U^{\dagger}_i(x_\bot-\hat{\imath})U^{(A)}_{i}(x_\bot-\hat{\imath})]\;. \nonumber 
\end{eqnarray}
To generate the boost non-invariant fluctuations we first generate the functions $f(\nu)$ and $e_i^a(p_\bot)$ in momentum space according to
\begin{eqnarray}
f(\nu)&=&e^{-b|\nu|}~\xi(\nu)~\sqrt{N_\eta a_\eta} \\
e_i^a(p_\bot)&=&g a~\Delta~\xi_i^a(p_\bot)~(N_\bot a) 
\end{eqnarray}
where $\xi(\nu)$ are uncorrelated Gaussian random numbers. After performing a Fourier transform to coordinate space, the fluctuations $\delta E_{\mu}^a(x_\bot,\eta)$ are calculated as
\begin{eqnarray}
\label{eq:dE1}
\delta E_i^a(x)&=&f'(\eta)~e_i^a(x_\bot)\;, \\
\delta E_{\eta}^a(x)&=&2\tau a_\tau^{-1}~f(\eta)~[f'(\eta)]^{-1}~\sum_{i=1,2}\mathcal{G}_{i}^a(x)\;,
\label{eq:dE2}
\end{eqnarray}
where $f'(\eta)$ is the lattice derivative of the function $f(\eta)$ according to
\begin{eqnarray}
f'(\eta)&=&a_\eta^{-1}[f(\eta)-f(\eta-\hat{\eta})]\;
\end{eqnarray}
and the Gauss constraint is implemented explicitly in Eqns.~(\ref{eq:dE1}) and (\ref{eq:dE2}).
\subsection{Lattice observables}
We will denote the diagonal components of the stress-energy tensor by the energy density $\epsilon(x)$, the transverse pressure $P_T(x)$ and the longitudinal pressure $P_L(x)$.
In terms of the electric and magnetic field strengths squared, as defined in Eq.~(\ref{eq:EandB}), they are given by
\begin{eqnarray}
\epsilon(x)&=&c_E\frac{\mathcal{E}_T^2}{2}+\frac{\mathcal{E}_L^2}{2}+c_B \frac{\mathcal{B}_T^2}{2}+\frac{\mathcal{B}_L^2}{2} \;, \\
P_T(x)&=&\frac{\mathcal{E}_L^2}{2}+\frac{\mathcal{B}_L^2}{2} \;, \\
P_L(x)&=&c_E\frac{\mathcal{E}_T^2}{2}-\frac{\mathcal{E}_L^2}{2}+c_B\frac{\mathcal{B}_T^2}{2}-\frac{\mathcal{B}_L^2}{2}\;.
\end{eqnarray}
These quantities are gauge invariant and satisfy the relation $\epsilon=2P_T+P_L$ at every position in space and time. In addition to the above quantities, we also study equal time correlation functions
\begin{eqnarray}
\Pi_L^2(\tau,\eta,\eta')&=&\left< P_L(\tau,x_\bot,\eta)P_L(\tau,y_\bot,\eta')\right>_T\;,
\end{eqnarray}
where $\langle.\rangle_T$ denotes ensemble average and average over transverse coordinates. We focus on the correlator in Fourier space with respect to relative rapidity, which is given by
\begin{eqnarray}
\Pi_L^2(\tau,\nu)&=&L_{\eta}^{-1}\int d\eta~d\eta'~\Pi_L(\tau,\eta,\eta')~e^{-i\nu(\eta-\eta')}\;
\end{eqnarray}
and usually show results for $\Pi_L(\nu)$, i.e. the square root of the above expression.

\end{document}